\documentclass[fleqn,usenatbib]{mnras}

\usepackage{newtxtext,newtxmath}

\usepackage[T1]{fontenc}

\usepackage[dvipsnames]{xcolor}

\DeclareRobustCommand{\VAN}[3]{#2}
\let\VANthebibliography\thebibliography
\def\thebibliography{\DeclareRobustCommand{\VAN}[3]{##3}\VANthebibliography}

\usepackage{graphicx}	
\usepackage{amsmath}	
\usepackage[normalem]{ulem}

\title[Parameter Inference on Accreting White Dwarfs]{Astrophysical Parameter Inference on Accreting White Dwarf Binaries using Gravitational Waves}

\author[S. Yi et al.]{
Sophia Yi,$^{1}$\thanks{E-mail: syi24@jh.edu }
Shu Yan Lau,$^{1}$
Kent Yagi$^{1}$
and Phil Arras$^{2}$
\\
$^{1}$Department of Physics, University of Virginia, Charlottesville, VA 22904, USA \\
$^{2}$Department of Astronomy, University of Virginia, Charlottesville, VA 22904, USA
}

\date{Accepted XXX. Received YYY; in original form ZZZ}

\pubyear{2023}

\begin{document}
\label{firstpage}
\pagerange{\pageref{firstpage}--\pageref{lastpage}}
\maketitle

\begin{abstract}
Accreting binary white dwarf systems are among the sources expected to emanate gravitational waves that the \textit{Laser Interferometer Space Antenna} (\textit{LISA}) will detect. We investigate how accurately the binary parameters may be measured from \textit{LISA} observations. We complement previous studies by performing our parameter estimation on binaries containing a low-mass donor with a thick, hydrogen-rich envelope. The evolution is followed from the early, pre-period minimum stage, in which the donor is non-degenerate, to a later, post-period minimum stage with a largely degenerate donor. We present expressions for the gravitational wave amplitude, frequency, and frequency derivative in terms of white dwarf parameters (masses, donor radius, etc.), where binary evolution is driven by gravitational wave radiation and accretion torques, and the donor radius and logarithmic change in radius ($\eta_{\rm d}$) due to mass loss are treated as model parameters. We then perform a Fisher analysis to reveal the accuracy of parameter measurements, using models from Modules for Experiments in Stellar Astrophysics (MESA) to estimate realistic fiducial values at which we evaluate the measurement errors.  We find that the donor radius can be measured relatively well with \textit{LISA} observations alone, while we can further measure the individual masses if we have an independent measurement of the luminosity distance from electromagnetic observations. When applied to the parameters of the recently-discovered white dwarf binary ZTF J0127+5258, our Fisher analysis suggests that we will be able to constrain the system's individual masses and donor radius using \textit{LISA}'s observations, given Zwicky Transient Facility’s (ZTF's) measurement of the luminosity distance.
\end{abstract}

\begin{keywords}
accretion, accretion discs -- gravitational waves -- binaries: close -- white dwarfs
\end{keywords}



\section{Introduction}

The first direct detection of gravitational waves (GWs) in 2015 came from the merger of a binary black hole~\citep{Abbott:2016blz}. Since then, the LIGO/Virgo Collaborations have additionally detected GW signals from numerous other binary black hole mergers, as well as several binary neutron star and neutron star-black hole mergers~\citep{TheLIGOScientific:2017qsa,LIGOScientific:2018mvr,LIGOScientific:2021qlt}. While LIGO and other ground-based detectors are able to detect GWs with frequencies from about 15 Hz to several kHz ~\citep{LIGOScientific:2018mvr}, the \textit{Laser Interferometer Space Antenna} (\textit{LISA}) is a space-based GW detector expected to launch in the mid-2030s with the ability to detect GWs in the frequency range of $\sim10^{-4}$ to $10^{-1}$Hz ~\citep{Audley:2017drz}. Among the astrophysical sources anticipated to emit GWs within this range are binary white dwarfs (WDs). In fact, for a 4-year observation period, some $\sim 10^4$ double white dwarfs (DWDs) are expected to be resolvable with \textit{LISA}~\citep{Lamberts:2019nyk}. For DWD systems emitting GWs at relatively high frequency, we anticipate being able to extract significant information from not only the GW frequency but also the GW frequency ``chirp,'' i.e., change in frequency over time ($\dot f$)~\citep{2012A&A...544A.153S}.  

Prospects of measuring astrophysical parameters of \emph{detached} binary WDs, in particular their individual masses, are studied in~\cite{2021MNRAS.500L..52W}. The authors employ universal relations between the tidal deformability and moment of inertia, and between the moment of inertia and WD mass, to express the finite-size effects entering in the gravitational waveform in terms of the individual masses. By conducting a Fisher analysis on this waveform expressed in terms of the masses, the authors show that \textit{LISA} will be able to measure individual masses of DWDs for sufficiently small binary separation and large stellar masses.

Accreting DWD systems with close separations can also be strong GW sources \citep{10.1111/j.1365-2966.2004.07479.x}. Some of these systems have been observed as cataclysmic variables and are identified as \textit{LISA} verification sources \citep{Stroeer_2006, Kupfer_18}. A few examples are V407 Vul \citep{Cropper_98}, HM Cancri \citep{refId0, 10.1046/j.1365-8711.2002.05471.x}, and ES Cet \citep{Warner_2002}. 
The population of these systems depends on the stability of the accretion. The influence of tidal synchronization on the formation of AM Canum Venaticorum (AM CVn) type binaries, a type of DWD system involving accretion of hydrogen-poor gas, has been studied in \cite{10.1111/j.1365-2966.2004.07564.x, 
 Gokhale_2007, Sepinsky_2014, 2015ApJ...806...76K}. They consider the stability criteria of the accretion, either through a disk or direct impact, taking into account the tidal synchronization torque.
 Population simulations of such accreting \textit{LISA} sources  have been performed in
\cite{2017ApJ...846...95K, Biscoveanu_2023}. In particular, the work by \cite{2017ApJ...846...95K} demonstrates that $\sim 10^3$ DWDs with negative chirp due to accretion may be observed by \textit{LISA}. In \cite{Biscoveanu_2023}, they employ a similar population study to further show that \textit{LISA} is able to constrain the tidal synchronization timescale ($\tau_0$) through its influence on the population distribution in the $f$-$\dot f$ parameter space.
In these studies, they only consider cold degenerate WDs. 
On the other hand,  accreting systems with an extremely low mass WD donor that has a thick hydrogen envelope are studied in \cite{Kaplan:2012hu}. These hydrogen-rich donors are what we would expect to see in an early, pre-period minimum stage of binary evolution. \cite{Kaplan:2012hu} highlights the importance of understanding the relative composition of hydrogen and helium in these WDs in order to infer the stability and behavior of the binary. These systems can be candidates of the observed inspiraling cataclysmic variables (e.g., HM Cancri) and may evolve into AM CVn binaries. 

In this paper, we investigate the possibility of directly measuring astrophysical parameters of \emph{accreting} DWDs given \textit{LISA}'s measurements of the amplitude ($A$), frequency ($f$), and frequency derivative ($\dot f$) of GWs emanated by the DWDs. We additionally build on the work of \cite{Kaplan:2012hu} by connecting the non-degenerate regime, in which donor WDs have a lingering hydrogen envelope, with the later, degenerate regime in which the zero-temperature mass-radius relation is valid. We study the evolution of accreting DWDs through the transition between these two regimes, as the orbital period of the binary first decreases due to gravitational radiation, then reaches a ``period minimum'' and begins to increase again as the effect of accretion takes over and drives the binary components apart again. Knowledge of the accretion physics for such DWDs allows us to parameterize their gravitational waveforms in terms of the individual masses and other parameters of interest. We then perform a Fisher analysis on this waveform to determine how well we can constrain the masses and other parameters given \textit{LISA}'s detections of GWs from accreting DWDs.

We find that with an independent measurement of the luminosity distance of our DWD systems from electromagnetic observations, we are likely to be able to measure the individual masses, donor radius, and exponent of the mass-radius relation given \textit{LISA}'s measurements of the GW amplitude, frequency, and frequency derivative. With \textit{LISA} observations alone, we lose our ability to constrain the individual masses, but we are still able to measure the other two parameters. 

The rest of the paper is organized as follows. In Sec.~\ref{waveform}, we introduce the parameterized gravitational waveform. In Sec.~\ref{sec:WD_properties}, we discuss how the mass-radius relations of our WDs differ in the degenerate versus non-degenerate regimes, introducing models of donors in the non-degenerate regime that we generate with a stellar evolutionary code. We also discuss the dynamical stability of accreting DWDs. Section~\ref{strain} illustrates the detectability of our DWD systems based on the relative magnitude of these systems' GW strain versus \textit{LISA}'s noise curve. Finally, our parameter estimation technique and results are given in Sec.~\ref{param estimation}, followed by discussion and conclusions in Sec.~\ref{sec:conclusion}.
The geometric units of $c=G=1$ are used in all of our equations, with the physical dimensions being recoverable through the conversion $1M_\odot = 1.5\mathrm{km} = 4.9\times 10^{-6}$s.

\section{Gravitational Waveform}\label{waveform}
 
The sky-averaged gravitational waveform, $h(t)$, for a DWD with donor mass $m_{\rm d}$ and accretor mass $m_{\rm a}$ is given by 
\begin{align}
h(t) = A\cos \phi(t).
\end{align}
In this expression, $A$ is the amplitude, given by~\citep{Cornish:2018dyw}
\begin{align}
\label{amplitude}
    A=\frac{8\mathcal{M}}{\sqrt{5}D}(\pi\mathcal{M}f)^{2/3}\,,
 \end{align} 
where $D$ is the luminosity distance and $\mathcal{M}$ is the chirp mass,
\begin{align}
    \mathcal{M} = \frac{(m_{\rm d}m_{\rm a})^{3/5}}{(m_{\rm d} + m_{\rm a})^{1/5}}.
\end{align}
Because the accretion processes we consider here occur on a timescale of Myr, the change in GW frequency is gradual enough for us to be able to neglect $\ddot f$ and higher derivatives, so that we can take the GW phase $\phi(t)$ to be given by~\citep{Shah:2014oea}
\begin{equation}
\label{phi}
    \phi(t)= \phi_0+2\pi f_0\delta t+\pi \dot{f}_0 \delta t^2,
\end{equation}
where the subscript 0 indicates the quantity measured at the initial time of observation, $t_0$, and $\delta t = t-t_0$. 

Examining Eqs.~\eqref{amplitude} and~\eqref{phi}, it is evident that in order to write the waveform in terms of our parameters of interest, we must express $f$ and $\dot f$ in terms of these parameters. 

We assume that the semi-major axis, $a$, adjusts itself during accretion such that the donor radius $r_{\rm d}$ always equals the Roche lobe radius, $r_{\rm L} a$, which we approximate with the fitting formula by~\cite{Eggleton:1983rx}:
\begin{align}
\label{rL}
    r_{\rm L}=\frac{0.49q^{2/3}}{0.6q^{2/3}+\ln(1+q^{1/3})},\quad q=\frac{m_{\rm d}}{m_{\rm a}}.
\end{align}
Taking the derivative of $r_{\rm d} = r_{\rm L} a$ leads us to a relation between the mass loss rate and the orbital separation:
\begin{equation}
\begin{aligned}
\label{dot a}
    \frac{\dot a}{a} =\frac{\dot m_{\rm d}}{m_{\rm d}}(\eta_{\rm d}-\eta_{\rm L}), 
\end{aligned}
\end{equation}
where $\eta_{\rm L}$ is the ratio between $\dot r_{\rm L}/r_{\rm L}$ and $\dot m_{\rm d}/m_{\rm d}$, given by 
\begin{equation}
\begin{aligned}
    \eta_{\rm L} & =\frac{d \ln r_{\rm L}}{d \ln m_{\rm d}} \\
   & =\Bigl[q(1-F)+1\Bigr]\frac{2(1+q^{1/3})\ln(1+q^{1/3})-q^{1/3}}{3(1+q^{1/3})\left[0.6q^{2/3}+\ln(1+q^{1/3})\right]},
\end{aligned}
\end{equation}
and $\eta_{\rm d}$ is the logarithmic change in radius due to mass loss,
\begin{align}
    \eta_{\rm d}=\frac{d \ln r_{\rm d}}{d \ln m_{\rm d}}.
\end{align} 
If $\eta_{\rm d}>0$, the donor shrinks as it loses mass; if $\eta_{\rm d}<0$, as is the case for a degenerate donor, the WD becomes larger in response to mass loss. As in \cite{Kaplan:2012hu}, we have introduced the mass-loss fraction, $F$, defined such that $\dot m_{\rm a}=-(1-F)\dot m_{\rm d}$, to indicate whether the mass transfer is conservative or not. When $F=0$, all mass lost by the donor is gained by the accretor, and there is no overall loss of mass from the binary; $F=1$ indicates that the accreted material is lost by the binary due to stellar winds, classical novae, etc. 

We then consider the angular momentum conservation. For circular binaries, the orbital angular momentum, $J$ is given by
\begin{align}
\label{ang_mom}
    J=m_{\rm d} m_{\rm a} \left( \frac{a}{M} \right)^{1/2},
\end{align}
with $M = m_{\rm a} + m_{\rm d}$. The conservation of this quantity is expressed as
\begin{align}
\label{ang_mom_cons}
\dot J = \dot J_\mathrm{gr} + \dot J_\mathrm{acc},
\end{align}
where $\dot J_\mathrm{gr}$ is the angular momentum carried away by gravitational radiation,
\begin{align}
\label{gr}
    \frac{\dot J_{\mathrm{gr}}}{J}=-\frac{32}{5}\frac{Mm_{\rm d}m_{\rm a}}{a^4},
\end{align}
and the ``accretion torque,'' $\dot J_{\mathrm{acc}}$, is given by
\begin{align}
\label{acc}
    \frac{\dot J_{\mathrm{acc}}}{J}=\frac{\dot m_{\rm d}}{m_{\rm d}}\sqrt{r_{\rm h}\left(1+q\right)}.
\end{align}
Here, $r_{\rm h}$ is the effective radius (in units of $a$) of material orbiting the accreting companion that carries the same amount of angular momentum as is lost due to the impact.

This accretion torque quantifies the angular momentum lost by the binary when accreted material impacts the companion star directly, rather than forming an accretion disk around the companion. We use a fitting formula for $r_{\rm h}$ that depends only on $q $~\citep{1988ApJ...332..193V} in the direct impact scenario, and set $r_{\rm h} = 0$ for the disk accretion case. To determine whether we have disk accretion or direct impact, we use Eq.~(6) of \cite{Nelemans_11} for the definition of the minimum radius of accreted material around the accreting WD,
\begin{equation}
\begin{aligned}
    \frac{r_{\rm{min}}}{a}&\approx 0.04948-0.03815 \,(\log_{10}q)\\
    &+0.04752\,(\log_{10}q)^2-0.006973 
 \,(\log_{10}q)^3\,,
\end{aligned}
\end{equation}
and assume disk accretion when the radius of the accretor, $r_{\rm a}$, is less than $r_{\mathrm{min}}$,  and direct impact for $r_{\rm a}>r_{\mathrm{min}}$. 

For the results shown in Sec.~\ref{param estimation}, the orbit was always wide enough for a disk to form, allowing us to neglect the torque term. This is because the lingering hydrogen envelopes in our models of donor WDs cause $r_{\rm d}$ (and therefore $r_{\rm{min}}$) to be relatively large, i.e., larger than $r_{\rm a}$. Nevertheless, at each step of our calculations, we determine whether $r_{\rm{min}}$ is larger or smaller than $r_{\rm a}$, and implement the appropriate equations for direct impact or disk-fed accretion correspondingly. 

Kepler’s third law is used to relate $f$, $M$, $r_{\rm d}$, and $ r_{\rm L} $:
\begin{align}
\label{fnew}
    f=\frac{1}{\pi}\sqrt{\frac{M}{(r_{\rm d}/r_{\rm L})^3}}\,,
\end{align}
where we have also used the fact that the GW frequency is twice the orbital frequency ($f=2f_{\rm{orb}}$). 
We find that in the degenerate regime, $f$ calculated according to Eq.~\eqref{fnew} depends almost entirely on $m_{\rm d}$, varying very little with different $m_{\rm a}$. This result agrees well with the findings of~\cite{Breivik:2017jip} in their analysis of DWDs with degenerate donors (see App.~\ref{fgw-md}).

\begin{figure*}
\includegraphics[width=0.95\textwidth]{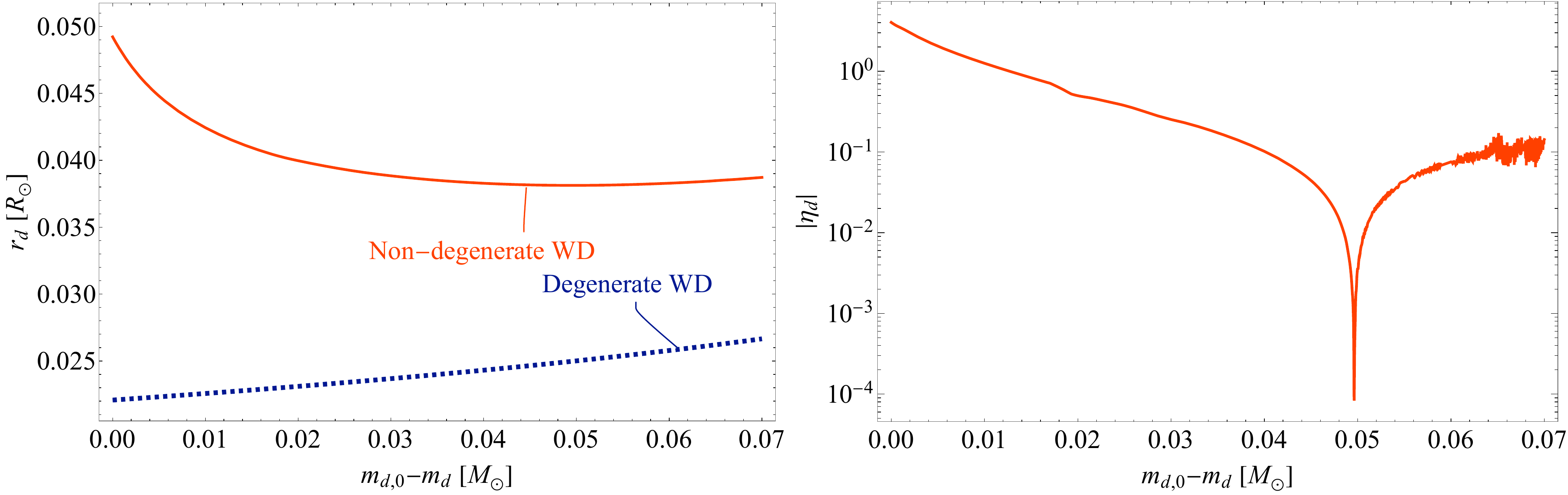} 
\caption{
\emph{Left:} Plot illustrating how the donor's radius as a function of stripped mass evolves differently between degenerate and non-degenerate regimes. As the donor loses mass, the degenerate WD's radius increases steadily ($r_{\rm d}\sim m_{\rm d}^{-1/3}$), whereas the non-degenerate WD's radius (calculated with MESA) first decreases before starting to increase as the donor becomes increasingly degenerate. As reported in~\citep{Kaplan:2012hu}, we find a remaining offset between the fully degenerate $r_{\rm d}(m_{\rm d})$ and the MESA model's radius even long after the hydrogen envelope has been stripped, which is a lingering effect of the non-degeneracy of the low-mass WD. The MESA model shown here was of a donor having a predominantly helium core of $0.153M_\odot$ and an initial hydrogen envelope of $0.006M_\odot$. \emph{Right:} $|\eta_{\rm d}|$ vs. stripped mass for the same MESA model of a donor WD. Numerical noise halted the model after about $0.07M_\odot$ of mass had been stripped from the donor. 
}
\label{fig:MESA}
\end{figure*}

Combining Eqs.~\eqref{dot a}, ~\eqref{ang_mom}, ~\eqref{ang_mom_cons},  ~\eqref{gr}, ~\eqref{acc}, and ~\eqref{fnew}, we now have a full expression for $\dot f$: 
\begin{equation}
\begin{aligned}
    \frac{\dot f}{f} &=-\frac{16}{5}\frac{M m_{\rm d} m_{\rm a}}{a^4}\times \\
    &\frac{\frac{F m_{\rm d}}{M}-3(\eta_{\rm d}-\eta_{\rm L})}{1+q(F-1)-\frac{F m_{\rm d}}{2M}-r_{\rm h}^{1/2}\left(1+q\right)^{1/2}+\frac{\eta_{\rm d}-\eta_{\rm L}}{2}},
    \end{aligned}
\label{fdot_total}
\end{equation}
i.e., $\dot f$ is a function of $m_{\rm d}$, $m_{\rm a}$, $r_{\rm d}$ (through $f$), $\eta_{\rm d}$, and $F$. We exclude the accretion torque term $-r_{\rm h}^{1/2}(1+q)^{1/2}$ originating from Eq.~\eqref{acc} whenever the orbit is wide enough for an accretion disk to form. Notice that one can take the limit of $\eta_{\rm d} - \eta_{\rm L} \to \infty$ in Eq.~\eqref{fdot_total} to find $\dot a/a$ and $\dot f/f$ without mass accretion. This is because, from Eq.~\eqref{dot a}, $\dot m_{\rm d}$ has to go to 0 when we set $\eta_{\rm d} - \eta_{\rm L} \to \infty$  while keeping $\dot a/a$ finite.
 
The difference from the expressions used in previous work, such as~\cite{Biscoveanu_2023} (other than the tidal coupling that we do not consider here), is that we (i) introduce the mass-loss fraction $F$ and (ii) further assume $r_{\rm d} = r_{\rm L} a$ at any time so that Eq.~\eqref{dot a} holds. 

With Eqs.~\eqref{amplitude}, ~\eqref{phi}, ~\eqref{fnew}, and~\eqref{fdot_total}, we have a gravitational waveform in terms of the six physical parameters $\phi_0,m_{\rm d},m_{\rm a},r_{\rm d},\eta_{\rm d}$ and $D$. We use this waveform in our Fisher analysis to determine the measurability of our parameters of interest. We note that despite there being only four raw model parameters in the waveform, $\theta^i=(\phi_0,A,f,\dot f)$, we can break some of the degeneracy between our six physical parameters by imposing priors on the individual masses (see Sec.~\ref{fisher}).


\section{Astrophysical Properties of Accreting Double White Dwarfs}
\label{sec:WD_properties}

We now use Modules for Experiments in Stellar Astrophysics \citep[MESA][]{Paxton2011, Paxton2013, Paxton2015, Paxton2018, Paxton2019} to study the mass-radius relation and $\eta_{\rm d}$ for WDs and consider the dynamical stability of mass transfer in accreting DWDs.

\subsection{Mass-radius Relations}\label{m-r}

The manner in which the donor WD responds to mass loss depends significantly on the composition of the donor. For fully degenerate WDs, we can use Eggleton's analytic formula to obtain $r_{\rm d}$ in terms of $m_{\rm d}$~\citep{1988ApJ...332..193V}. In this cold temperature regime, $r_{\rm d}$ goes roughly as $m_{\rm d}^{-1/3}$, so $\eta_{\rm d}\sim-1/3$ for a range of $m_{\rm d}$ values.

If the donor is not fully degenerate, $r_{\rm d}$ does not vary with $m_{\rm d}$ in such a simple manner. To reach this conclusion, we used MESA to model mass loss from dozens of WDs containing a range of core and envelope masses. The radius and $\eta_{\rm d}$ of one model are shown in Fig.~\ref{fig:MESA}. To construct this model, we used MESA to 
evolve a $M=$1.5$M_\odot$ pre-main sequence star to the red giant branch, and stopped the evolution when the helium core had mass $0.153M_\odot$.
The hydrogen-rich envelope was then rapidly removed until the envelope was reduced to $0.006M_\odot$. In Fig.~\ref{fig:MESA}, $|\eta_{\rm d}|$ begins at this point of the simulation; the initial positive value of $\eta_{\rm d}$ reflects the lingering hydrogen envelope surrounding the donor WD. We then simulate mass loss from the donor, causing the WD to become increasingly degenerate as hydrogen is transferred away, resulting in a decreasing $\eta_{\rm d}$ function. Eventually, $\eta_{\rm d}$ passes through zero (seen in the cusp around $0.052M_\odot$ of stripped mass). After this point, the donor increases in size as it continues losing mass, causing the orbital separation of the binary to increase\footnote{The MESA model halted after $\sim0.07M_\odot$ due to significant numerical noise, likely due to insufficient resolution near the outer boundary. The plot in Fig.~\ref{fig:MESA} shows one of the longest-lasting models we were able to obtain.}.

The reader may observe that even at the far right end of the left panel of Fig.~\ref{fig:MESA}, there is a remaining vertical offset between values of the degenerate and non-degenerate radii. This behavior is also reported by~\cite{Kaplan:2012hu}, where the authors note that if the WD had lower hydrogen content and enough time to cool, we would expect the degenerate and non-degenerate radii to eventually converge.

In the sections that follow, we will use the MESA model described here to obtain realistic fiducial values at which to evaluate the error on the astrophysical parameters, $r_{\rm d}$ and $\eta_{\rm d}$. In so doing, we account for the fact that the accreting DWDs that \textit{LISA} observes may have low-mass donors with lingering hydrogen envelopes, causing the radius (and $\eta_{\rm d}$) to be larger than what a fully degenerate WD would have. In other words, using the MESA model for fiducial values of $r_{\rm d}$ and $\eta_{\rm d}$ allows us to apply our parameter inference to DWDs in a near-period minimum stage of evolution, when the GW strain is likely to be highest (as we will show in Sec.~\ref{strain}). 

\begin{figure}
\centering\includegraphics[width=0.45\textwidth]{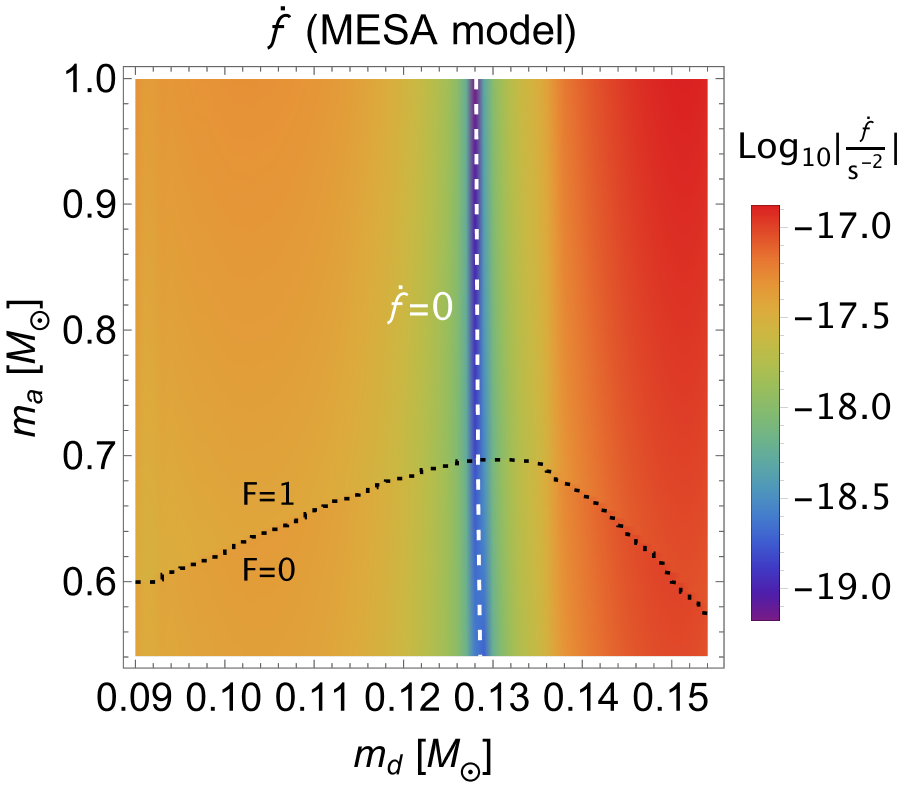} 
\caption{The rate of change of GW frequency, $\dot f$, for a DWD containing a donor modeled by MESA (mass-radius relation plotted in Fig.~\ref{fig:MESA}). Going from right to left on the plot, $\dot f$ goes from strictly positive to zero at the dotted white line, to strictly negative.}
\label{fig:fdot}
\end{figure}

\begin{figure*}
\centering\includegraphics[width=0.530\textwidth]{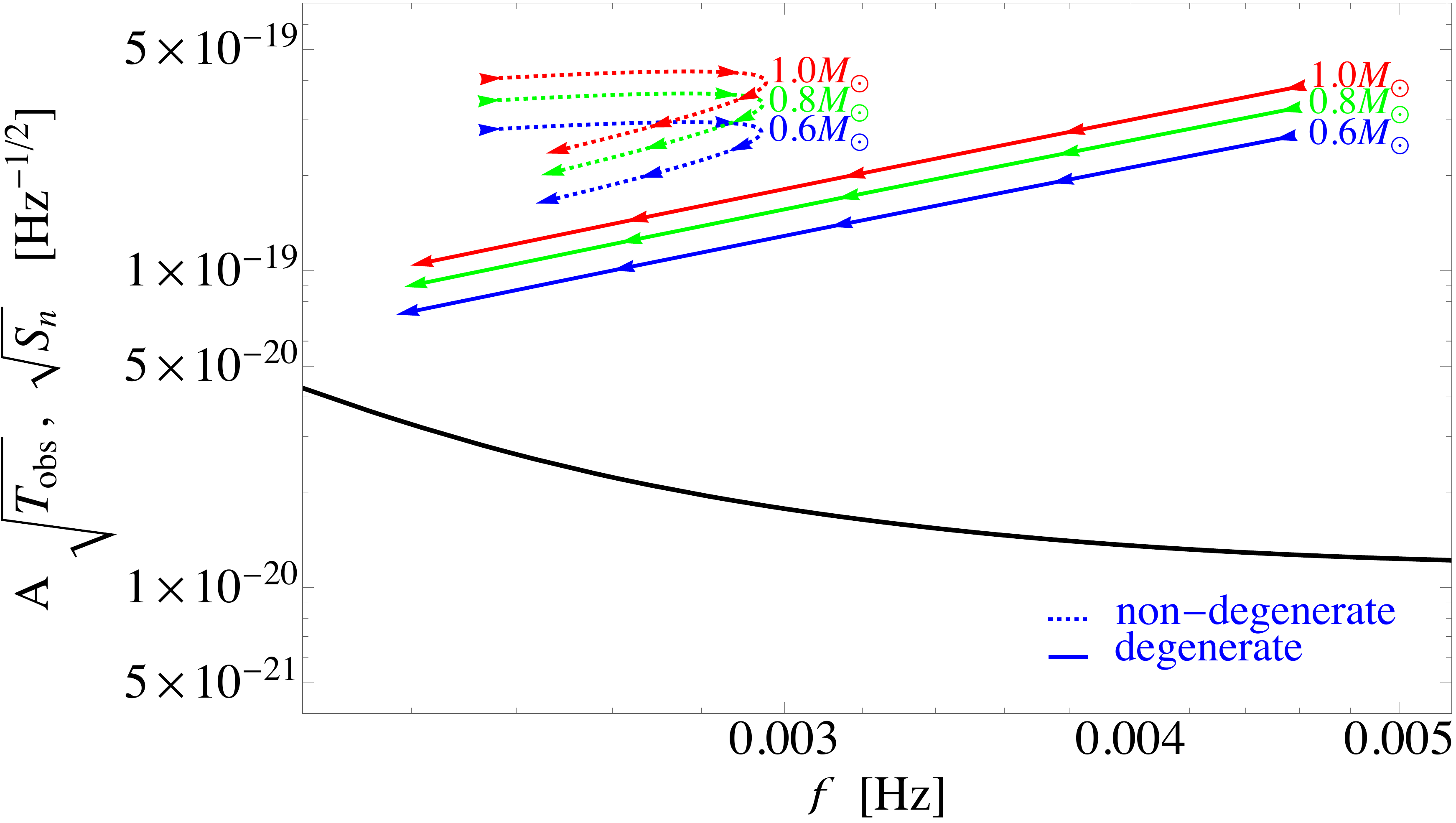} 
\centering\includegraphics[width=0.355\textwidth]{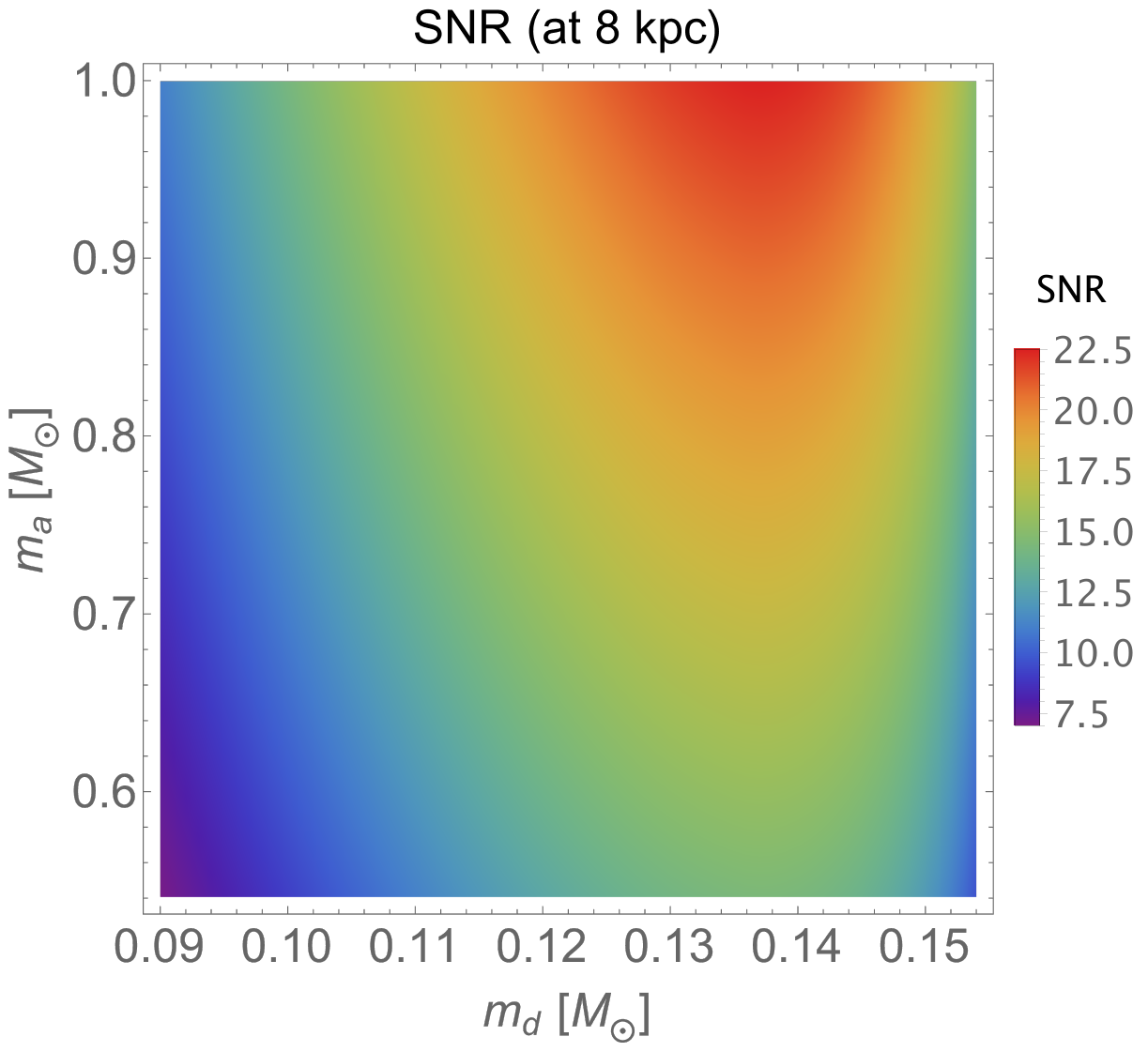} 
\caption{
\emph{Left:} GW strain $[A\times(\mathrm{T_{obs}})^{1/2}$; (1.0, 0.8, 0.6)~M$_\odot$] and \textit{LISA}'s noise curve (solid curve at the bottom) vs. GW frequency for degenerate (solid) and non-degenerate (dashed) DWD systems for a variety of initial accretor masses at a luminosity distance of 8kpc. Arrows show the direction of evolution. The frequency ranges correspond to initial and final donor   
masses $(m_\mathrm{initial},m_\mathrm{final})=(0.090,0.040)M_\odot$ and $(m_\mathrm{initial},m_\mathrm{final})=(0.154,0.085)M_\odot$ for the solid and dashed lines, respectively. 
Note that the evolution tracks span a timescale of $\sim$Myr based on Eq.~\eqref{fdot_total}, which is much longer than the observation time of \textit{LISA}. \emph{Right:} The signal-to-noise ratio (SNR) computed for a DWD system at 8 kpc with a donor modeled by MESA. Based on these plots, we expect GWs from the DWDs we study to be detectable at distances around 8kpc. 
} 
\label{fig:strain}
\end{figure*}

We note that based on Eq.~\eqref{dot a}, with $\dot{m}_d<0$ (donor losing mass), orbital separation will eventually start to grow ($\dot{a}>0$) as $\eta_{\rm d}$ decreases below $\eta_{\rm L}$. In a similar manner, using Eqs.~\eqref{dot a} and~\eqref{fdot_total}, we see that $\dot f$ will also switch from positive to negative as the magnitude of $\eta_{\rm d}$ decreases in comparison with $\eta_{\rm L}$. Figure~\ref{fig:fdot} shows this change of $\dot f$ for a binary with a donor mass described by the MESA model in Fig.~\ref{fig:MESA}; time evolves from right to left on this plot. All $\dot f$ values on the right of the dashed white line (period minimum, where $\dot f=0$) are positive, and all values to the left are negative. In this plot, we also see a discontinuity in the region between $m_{\rm a} = 0.6 M_\odot$ and $m_{\rm a} = 0.7M_\odot$ (marked by the dotted black line). This is due to our choice of a discontinuous transition of $F$ between a region of parameter space in which $F$ equals 0 (lower portion of the plot) to a region in which $F=1$ (upper portion). We will see shortly that our choice of $F$ depends on the magnitude of the accretion rate.

\subsection{Dynamical Stability for the DWD systems} \label{stability}

The mass transfer process for the DWDs considered here is expected to be unstable for certain mass ratios. Such unstable mass transfer causes dynamical instability of the binary, and since these unstable binaries will be short-lived, we are less likely to observe them within a 4- or even 10-year \textit{LISA} observation time. We follow~\cite{1982ApJ...254..616R} in taking mass transfer to be stable if we have a self-consistent solution in Eqs.~\eqref{dot a} and \eqref{fdot_total} assuming $\dot m_{\rm d} < 0$. Revisiting Eqs.~\eqref{dot a} and~\eqref{fdot_total}, we arrive at the following criterion for the dynamical stability of the binary: 
\begin{equation}
    1+q(F-1)-\frac{Fm_{\rm d}}{2M}-r_{\rm h}^{1/2}\left(1+q\right)^{1/2}
    +\frac{\eta_{\rm d}-\eta_{\rm L}}{2}>0.
\label{criterion}
\end{equation}

From the above expression, it is evident that dynamical stability is dependent on the value of the mass-loss fraction $F$, which is determined by whether the accreted material can be burned stably. We take the criterion for stable hydrogen burning from~\cite{Kaplan:2012hu} and adopt 

\begin{equation}
F=
    \begin{cases}
1 \quad (|\dot m_{\rm d}| < \dot m_\mathrm{c})\,, \\ 
0 \quad (|\dot m_{\rm d}| > \dot m_\mathrm{c})\,,
    \end{cases}
\end{equation}
where the critical mass loss rate is given by~\cite{Kaplan:2012hu},
\begin{equation}
    \dot m_\mathrm{c} = 10^{-7}\left(\frac{m_{\rm a}}{M_\odot} - 0.5357\right)M_\odot \, \mathrm{yr}^{-1}\,,
\end{equation}
which takes into account the reduced metallicity of the accreting WD.

As mentioned previously, for the results shown in Sec.~\ref{param estimation}, the orbit was always wide enough for an accretion disk to form. As a result, the accretion torque term ($-r_{\rm h}^{1/2}(1+q)^{1/2}$) in Eq.~\eqref{criterion} was always 0 for the calculations shown here, leading to greater dynamical stability (i.e., more regions in which the left-hand side is positive). This is a direct consequence of the hydrogen-rich nature of the donor WD; for a degenerate donor of the same mass, we often find $r_{\rm a}>r_{\rm{min}}$, in which case the accretion torque term would be nonzero. 
 In this case, this extra torque gives a larger $|\dot f|$ and might reduce the range of $q$ for stable mass transfer \citep{10.1111/j.1365-2966.2004.07564.x}.

As mentioned in~\cite{Kaplan:2012hu}, the hydrogen-rich nature of a low-mass donor also enhances the likelihood of dynamical stability directly, since such a WD has a larger positive $\eta_{\rm d}$, which leads to a more positive left-hand side of Eq.~\eqref{criterion}.

Lastly, we note that~\cite{Kaplan:2012hu} additionally account for the possibility of unstable helium burning at higher mass transfer rates, in which case we could again have $F>0$. It would be interesting to implement a more careful analysis of the different scenarios in which the binaries discussed here might undergo nonconservative mass transfer.

\section{GW strain vs. \textit{LISA}'s Noise Curve}\label{strain}
The left panel of Fig.~\ref{fig:strain} presents the GW strain compared to \textit{LISA}'s noise curve over a range of GW frequencies. 
The dashed curves model the GW strain during an earlier stage of DWD evolution (near-period minimum), when the donor has some amount of lingering hydrogen. We construct these curves using Eqs.~\eqref{amplitude} and~\eqref{fnew}, along with the mass-radius relation from the MESA model shown in Fig.~\ref{fig:MESA}. 
The solid lines show the GW strain at a much later stage of evolution (post-period minimum), when the donor is fully degenerate and $f$ is steadily decreasing toward the bottom left-hand corner of the plot. These solid curves are constructed using Eggleton's cold-temperature radius-mass formula~\citep{1988ApJ...332..193V}. 

We see that in both the early (dashed) and late (solid) stages of evolution of these binaries, the GW strain (plotted as $A\times\mathrm{T_{obs}}^{1/2}$, where $\mathrm{T_{obs}}$ is the observation time that we take to be 4 years) is up to one order of magnitude higher than \textit{LISA}'s noise curve ($S_n(f)$;~\cite{Cornish:2018dyw}). The right panel of Fig.~\ref{fig:strain} shows the signal-to-noise ratio (SNR) at a luminosity distance of 8kpc. If we follow~\cite{2017ApJ...846...95K} in taking SNR=5 as the minimum SNR for detectability, Fig.~\ref{fig:strain} confirms that \textit{LISA} should be able to detect the DWDs discussed here. We note that the GW strain is higher for the non-degenerate case because at the same luminosity distance, a binary with a non-degenerate donor must have a larger donor mass, and therefore chirp mass, to emanate GWs at the same frequency as a comparable binary with a degenerate donor. The larger chirp mass at identical $D$ and $f$ leads to a larger $A$ (see Eq.~\eqref{amplitude}).

Finally, we note that our calculations for degenerate DWDs in the left panel of Fig.~\ref{fig:strain} agree well with the SNR results in Fig. 6 of ~\cite{2017ApJ...846...95K}. Like them, at D=8kpc and for the mass ranges we consider ($m_{\rm d}\simeq0.09-0.15M_\odot, m_{\rm a}\simeq0.54-1.00M_\odot$), we also have a majority of parameter space with SNR between 5 and 10, as well as smaller regions with SNR $<5$ and $\geq10$.

\section{Astrophysical Parameter Inference}\label{param estimation}

Let us now move on to carrying out our parameter estimation for the accreting DWDs with \textit{LISA}. We first explain our methodology and next present our findings.

\subsection{Fisher Method}\label{fisher}

Given our gravitational waveform derived in Sec.~\ref{waveform}, we can estimate the statistical error on parameters due to the detector noise using a Fisher information matrix (FIM)~\citep{cutler1998,2012A&A...544A.153S,Shah:2014oea}. This method of parameter estimation assumes stationary and Gaussian detector noise. 

The FIM is defined as 
\begin{equation}
\label{Gamma}
\Gamma_{ij} = \left( \frac{\partial h}{\partial \theta^i} \bigg|\frac{\partial h}{\partial \theta^j} \right)\,,
\end{equation}
where the partial derivatives of the waveform $h$ are taken with respect to the parameters of interest described in the previous section, 
\begin{equation}
\label{eq:parameter1}
\theta^i = (\phi_0,m_{\rm d},m_{\rm a},r_{\rm d},\eta_{\rm d},D).
\end{equation}
The inner product in Eq.~\eqref{Gamma} is given by 
\begin{equation}\label{eq:inner}
(a|b) = 4 \int^\infty_0 \frac{\tilde a^*(f) \tilde b(f)}{S_n(f)}df \approx \frac{2}{S_n(f_0)} \int^T_0 a(t) b(t) dt\,,
\end{equation}
with spectral noise density $S_n$ and observation time $T$. Tildes indicate Fourier components, and the asterisk denotes the complex conjugate of $\tilde a(f)$. We take \textit{LISA}'s $S_n$ from~\citet{Cornish:2018dyw}. The monochromatic nature of DWD signals is assumed in our approximation, $S_n(f) \approx S_n(f_0)$, and we use Parseval's theorem to convert the inner product defined in the frequency domain to an integral in the time domain. 
By inverting the FIM defined in Eq.~\eqref{Gamma}, we obtain the 1-$\sigma$ uncertainty on each of the parameters: 
\begin{align}
\Delta \theta^i = \sqrt{(\Gamma^{-1}){}_{ii}}.
\end{align}

\begin{figure*}
\centering\includegraphics[width=0.45\textwidth]{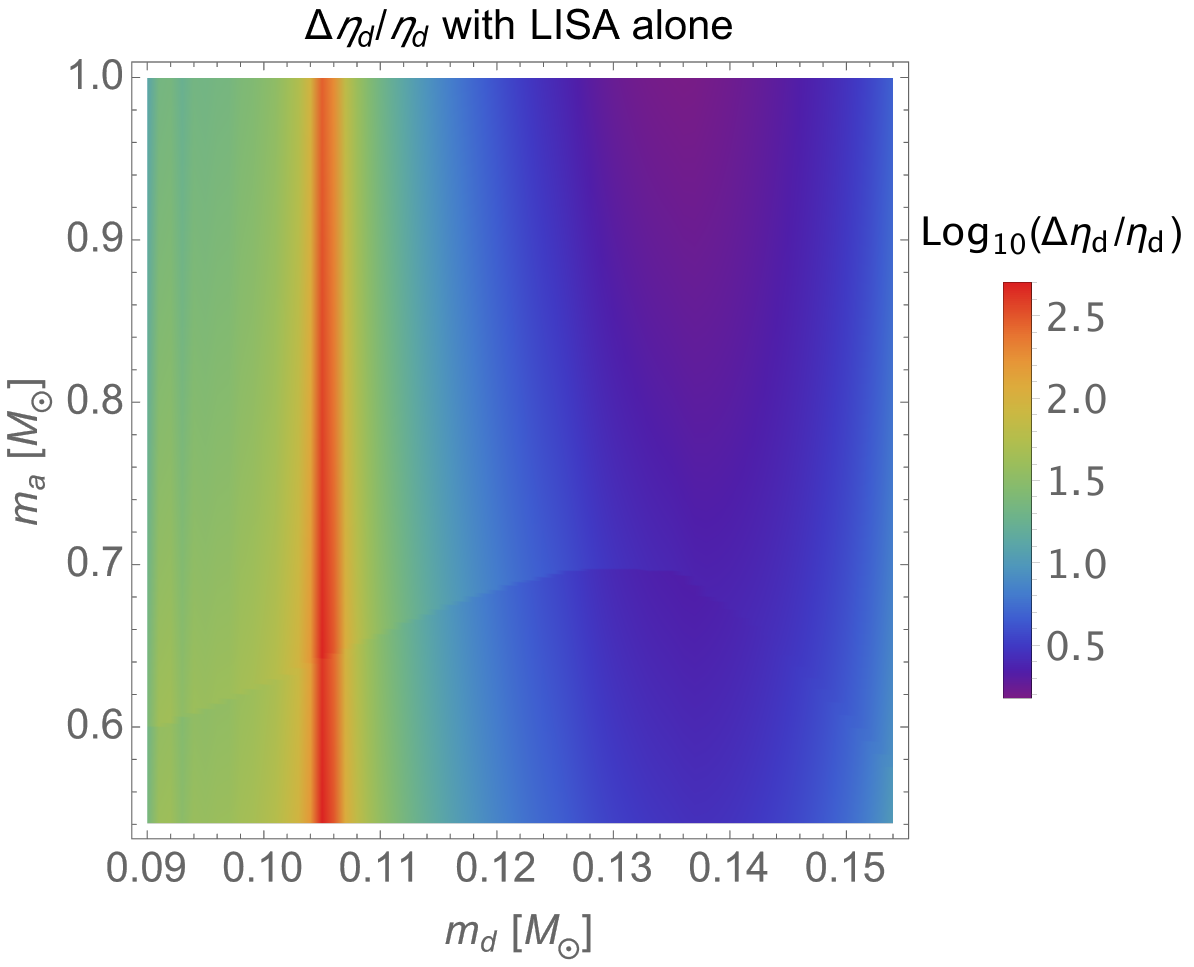} 
\centering\includegraphics[width=.45\textwidth]{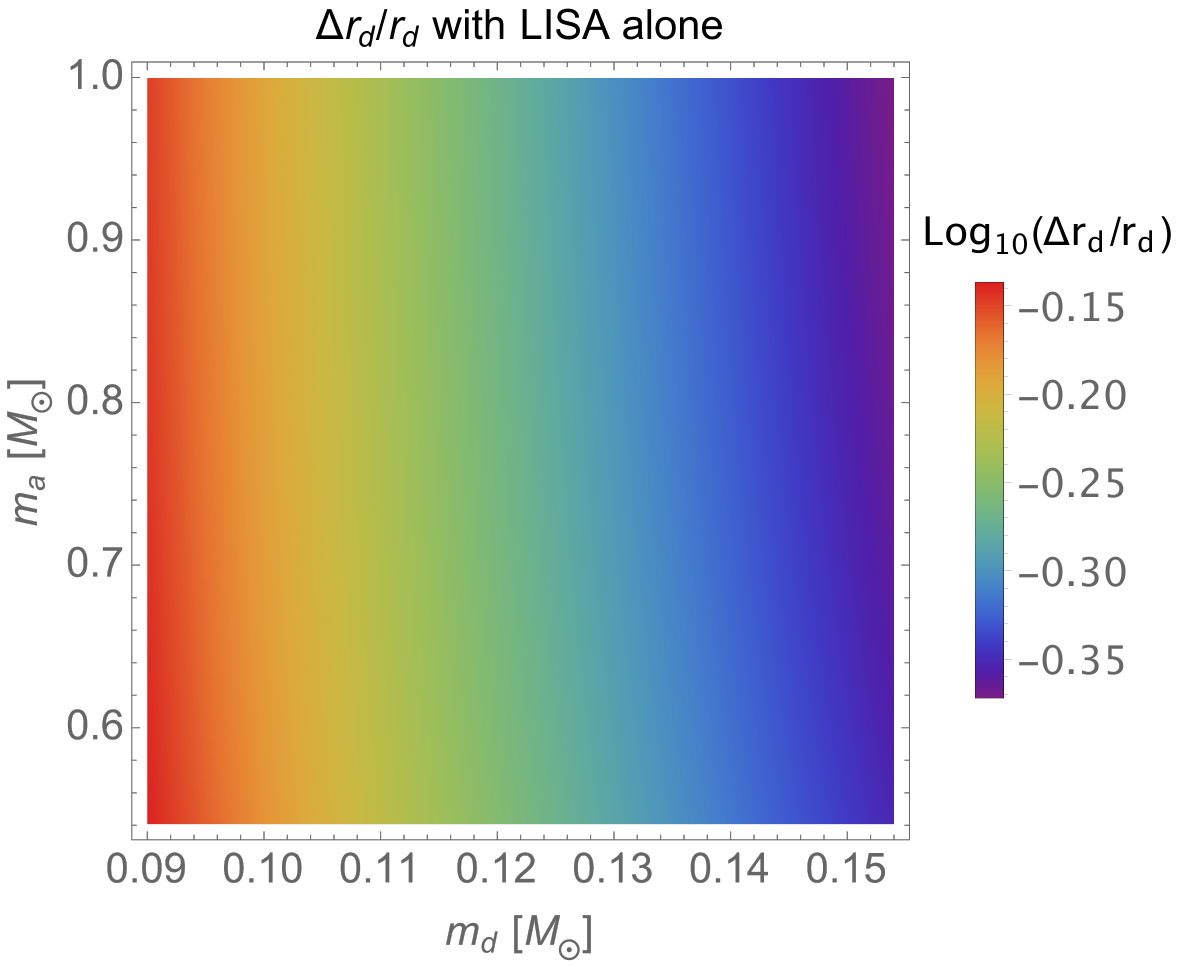} 
\centering\includegraphics[width=0.45\textwidth]{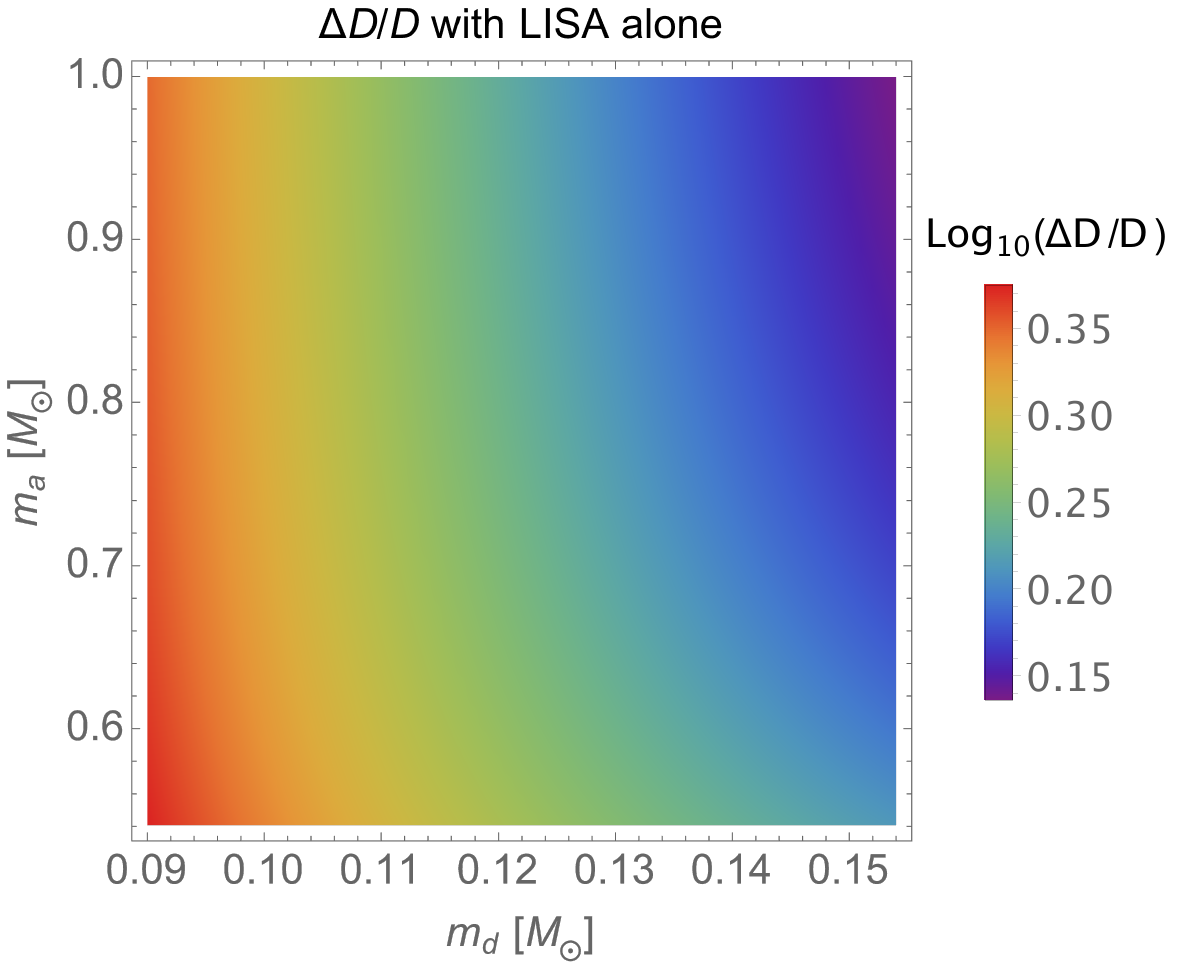} 
\caption{Fractional error on $\eta_{\rm d}$, $r_{\rm d}$, and $D$ for non-degenerate DWD systems, calculated via our Fisher analysis for \textit{LISA} observations alone. Since inverting the FIM merely returns the priors on the individual masses, giving no new constraints on the mass values, we do not show plots for these parameters. Fiducial values for $\eta_{\rm d}$ and $r_{\rm d}$ were obtained from the MESA model shown in Fig.~\ref{fig:MESA}. 
For the range of parameters shown here, the orbit was wide enough for an accretion disk to form, causing the accretion torque term $-r_{\rm h}^{1/2}(1+q)^{1/2}$ to be zero in Eqs.~\eqref{fdot_total} and~\eqref{criterion}.
}
\label{fig:6_params}
\end{figure*}

We further impose Gaussian priors on $m_{\rm d}$ and $m_{\rm a}$, with the priors $\sigma_{\theta^i}$ defined such that~\citep{Poisson:1995ef,cutlerflanagan,Carson:2019kkh}
\begin{align}
\Delta \theta^i = \sqrt{(\tilde \Gamma^{-1}){}_{ii}}\,, \quad \tilde \Gamma_{ij} = \Gamma_{ij}+\frac{1}{\sigma_{\theta^i}^2} \delta_{ij}.
\end{align}
As previous studies have found that shell flashes occur in the hydrogen envelope of donors with $m_{\rm d}\gtrsim0.2M_\odot$~\citep{Althaus2001,Panei2007}, we set the prior on the donor to $\sigma_{m_{\rm d}}=0.2M_\odot$. Requiring the accretor WD to have a larger mass than the donor WD, we set the prior on the accretor to $\sigma_{m_{\rm a}}=0.8M_\odot$, as WDs with masses much higher than $\sim1.0M_\odot$ are less common.

For fiducial values, we take $\phi_0=3.666$ rad and $D=8$kpc unless otherwise stated, and vary $(m_{\rm d},m_{\rm a})$. The MESA model in Fig.~\ref{fig:MESA} is used for the fiducial values of $r_{\rm d}(m_{\rm d})$ and $\eta_{\rm d}(m_{\rm d})$. Our results are shown for an observation time of $\mathrm{T_{obs}}=4$ years. 

\subsection{Results: Gravitational-wave Observations Alone}\label{no GAIA}

We begin by presenting results with \textit{LISA} observations alone. Our parameter set is
\begin{equation}
    \theta^i=(\phi_0,m_{\rm d},m_{\rm a},r_{\rm d},\eta_{\rm d},D)\,.
\end{equation}
The error on the parameters $\eta_{\rm d}, r_{\rm d},$ and $D$ are given in Fig.~\ref{fig:6_params}. Unfortunately, in this case, we find that our Fisher analysis merely returns the priors we impose on the masses, i.e., we gain no additional constraints on the individual masses of the DWDs.

Although we are unable to constrain the individual masses with \textit{LISA} observations alone, there are large regions of parameter space in which the fractional error on $r_{\rm d}$ is smaller than the measurability threshold, $\Delta r_{\rm d}/r_{\rm d}=1$. The same cannot be said for $D$ or $\eta_{\rm d}$; our Fisher analysis suggests that \textit{LISA} cannot determine these parameters for binaries in consideration. 

In the plot of $\Delta \eta_{\rm d}/\eta_{\rm d}$, we see the same discontinuity due to switching $F$ between 0 and 1 that we saw in Fig.~\ref{fig:fdot}. The fractional error on $\eta_{\rm d}$ is very large on the left side of the plot, which is partially due to the smallness of the parameter itself (see right panel of Fig.~\ref{fig:MESA}). In particular, there is a peak in the fractional error near $m_{\rm d}=0.105M_\odot$, corresponding to where $\eta_{\rm d}$ crosses zero, causing $\Delta \eta_{\rm d}/\eta_{\rm d}$ to be very large. However, even all the way to the right of the plot, where $\eta_{\rm d}>1$, the fractional error is generally greater than one, suggesting that we will be unable to constrain $\eta_{\rm d}$ from \textit{LISA}'s observations.

Let us comment on how the measurement errors on $r_{\rm d}$, $\eta_{\rm d}$, and $D$ scale with $f$, $\dot f$ and $A$. We derive the scaling by studying the measurement errors without correlations between parameters\footnote{ Error on $\theta^i$ without correlation is obtained as
\begin{equation}
\Delta \theta^i = 1/\sqrt{\Gamma_{ii}}\,.    
\end{equation}
}. First, we find that the fractional error on $r_{\rm d}$ scales inversely with the signal-to-noise ratio (SNR) times $(df/dr_{\rm d})\times r_{\rm d}$. This is intuitive; we see from Eq.~\eqref{fnew} that $f$ depends significantly on $r_{\rm d}$ through the orbital separation, and the error should of course decrease with a larger SNR. The extra factor of $r_{\rm d}$ accounts for the fact that we compare the derivative against our plots of fractional error (i.e., $\Delta r_{\rm d}$ times a factor of $1/r_{\rm d}$). In a similar manner, we find that the error on $\eta_{\rm d}$ scales inversely with SNR$\times 
(d\dot f/d\eta_{\rm d})\times\eta_{\rm d}$, which is sensible, as $\eta_{\rm d}$ only appears in $\dot{f}$ and not $A$ or $f$. Finally, we find that the error on $D$ scales
inversely with SNR $\times (dA/dD)\times D$. This is also as we expect; the only place luminosity distance appears in our parameterized gravitational waveform is through the amplitude (Eq.~\eqref{amplitude}). We note that the clear dependence of $\eta_{\rm d}$, $r_{\rm d}$, and $D$ on $\dot f$, $f$, and $A$, respectively, explains the appearance of the discontinuity in the plot of $\eta_{\rm d}$ alone: the mass loss fraction, $F$, only enters the waveform through $\dot f$, so the change between $F=0$ and $F=1$ does not alter error calculations on $r_{\rm d}$ and $D$. For more plots and discussion on the scaling of parameter error with various derivatives of the waveform, see App.~\ref{scaling}.

To summarize, in the absence of an electromagnetic counterpart to \textit{LISA}'s measurements of accreting DWDs, our Fisher analysis suggests that we will only likely be able to constrain $r_{\rm d}$ out of the six parameters appearing in our gravitational waveform. 

\subsection{Results: Gravitational-wave Observations with Electromagnetic Counterparts}\label{GAIA}

Let us now consider the case where we have electromagnetic (EM) counterparts. A recent paper has shown that at least $\sim60$ DWDs with helium-rich donors are expected to be observable by both \textit{LISA} and Gaia~\citep{Breivik:2017jip}. For these DWD systems, we can obtain an independent 
constraint of the luminosity distance $D$ from Gaia. Based on previous EM observations of the luminosity distance of compact binaries, we may expect to constrain $D$ to an error $\mathcal{O}(10\%)$ with an EM counterpart to \textit{LISA}~\citep{Burdge:2023mtw, Ramsay:2004tw}. We therefore investigate how our results change if we add a Gaussian prior on the luminosity distance, treating the parameter as partially constrained by an EM counterpart to \textit{LISA}'s observations.
\begin{figure*}
\centering\includegraphics[width=0.49\textwidth]{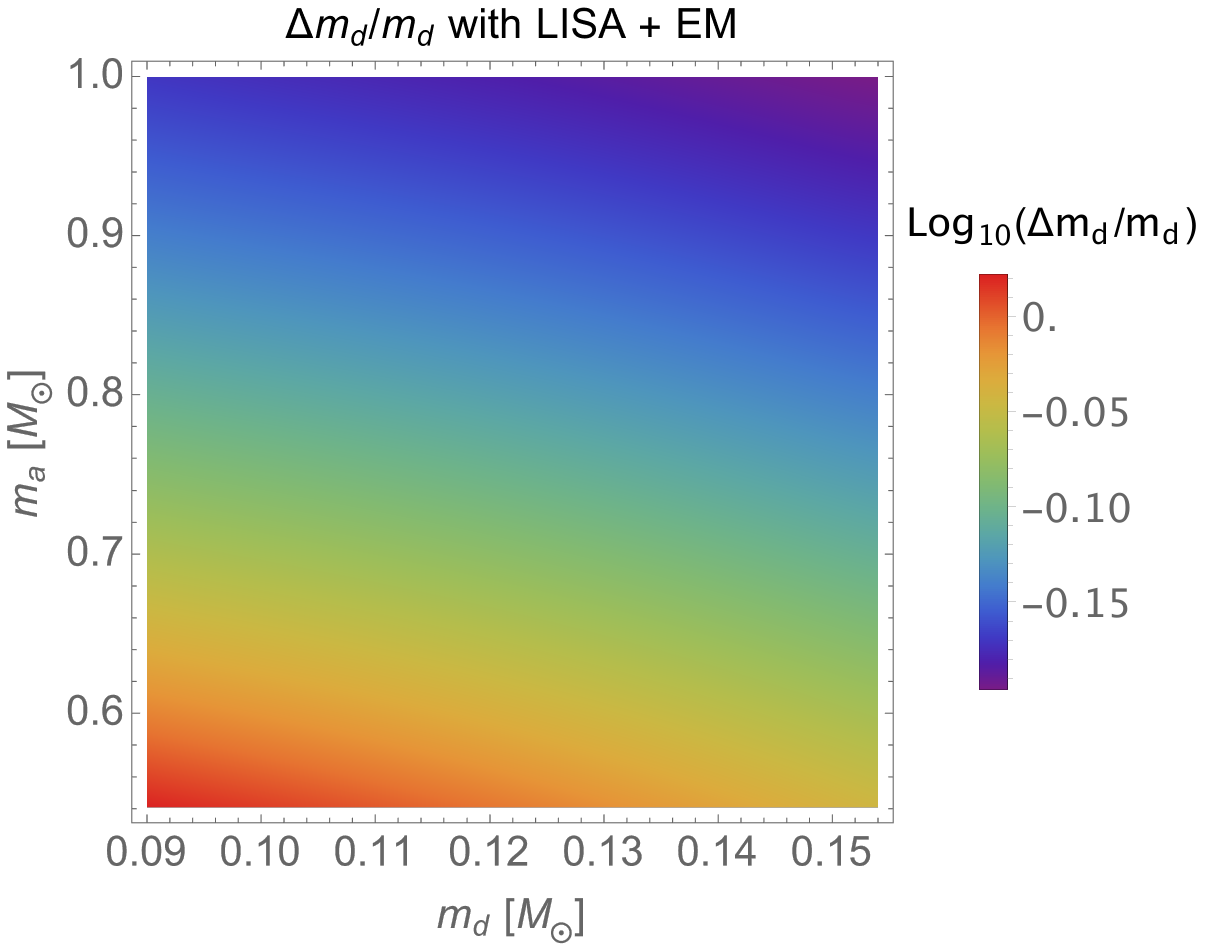} 
\centering\includegraphics[width=.48\textwidth]{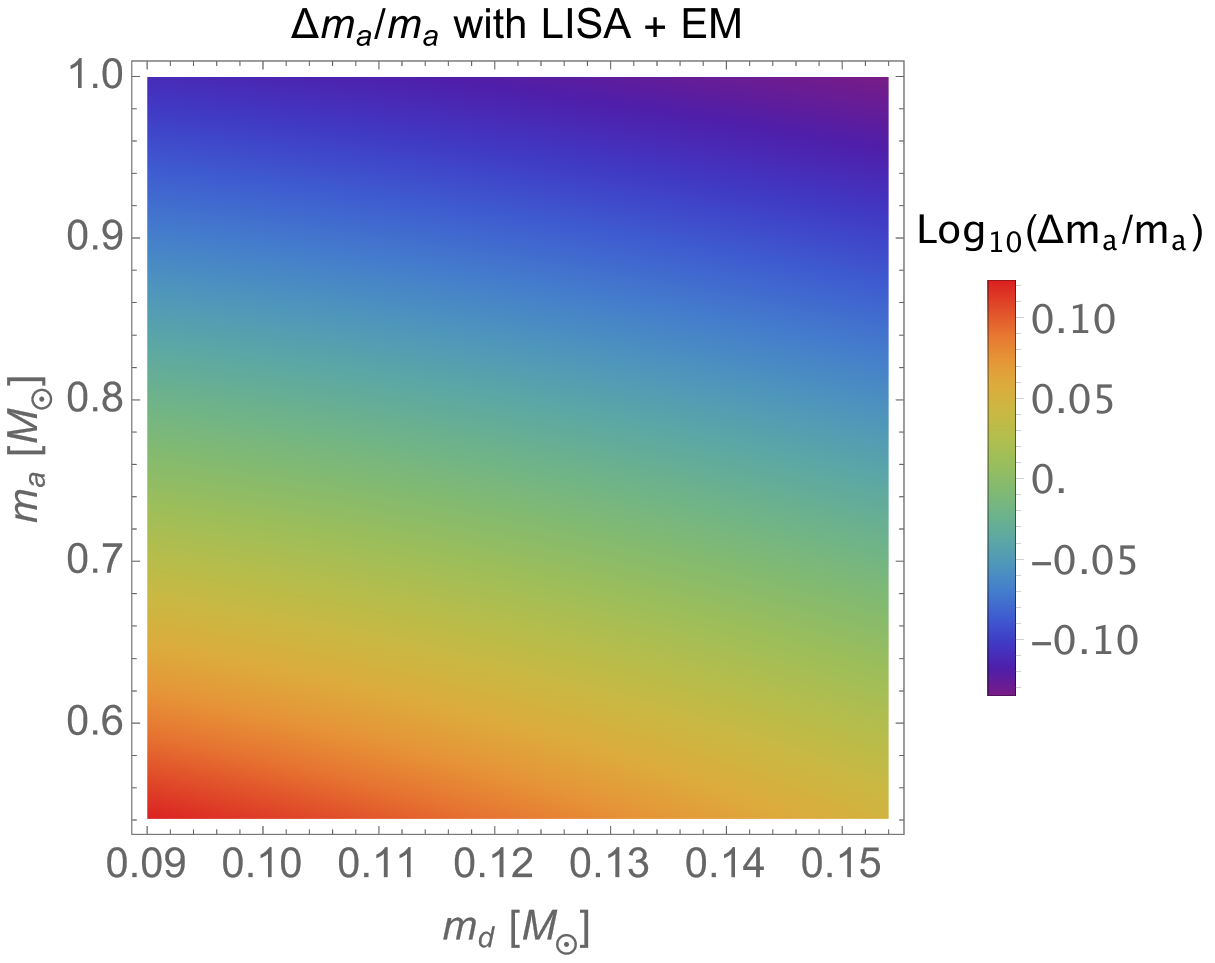} 
\centering\includegraphics[width=0.49\textwidth]{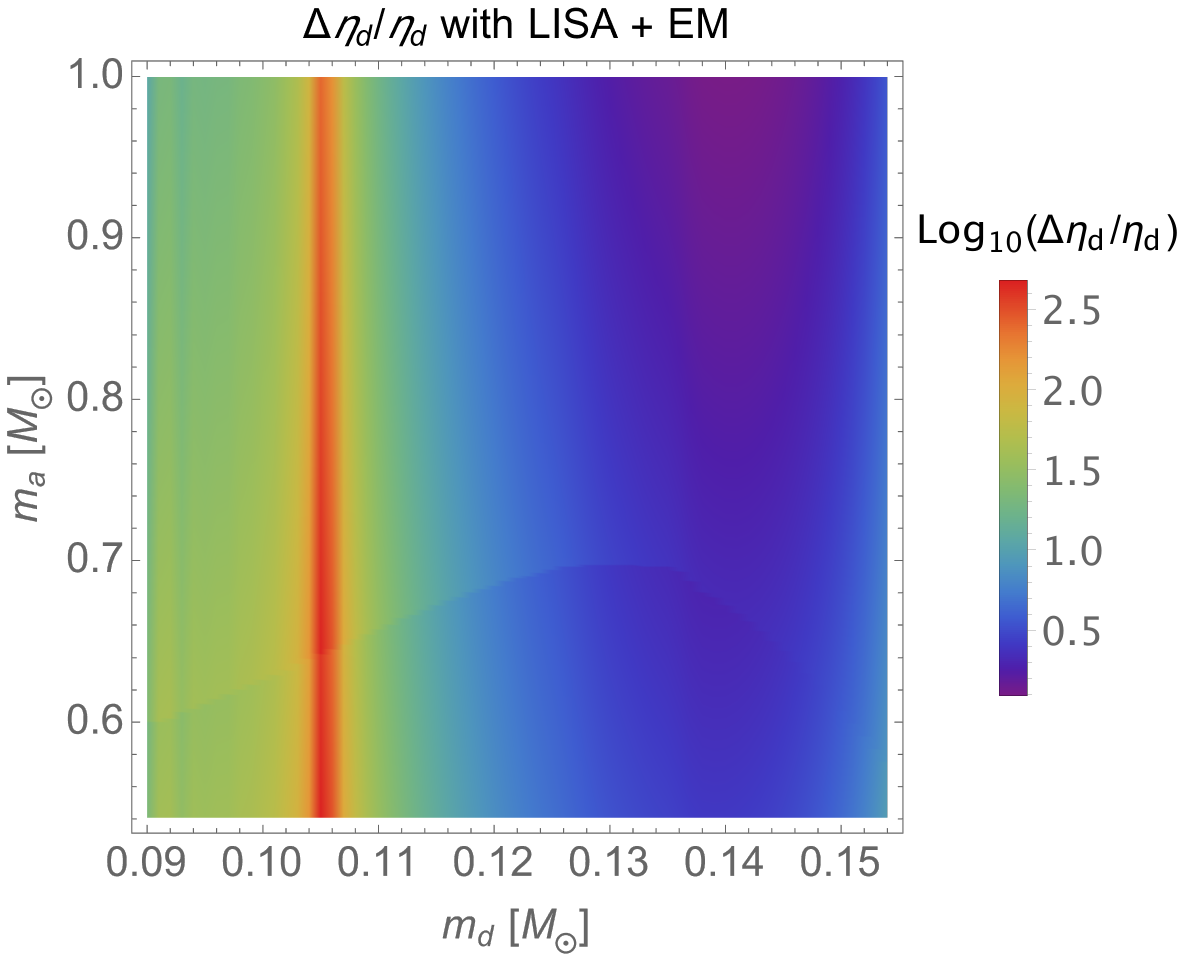} 
\centering\includegraphics[width=.48\textwidth]{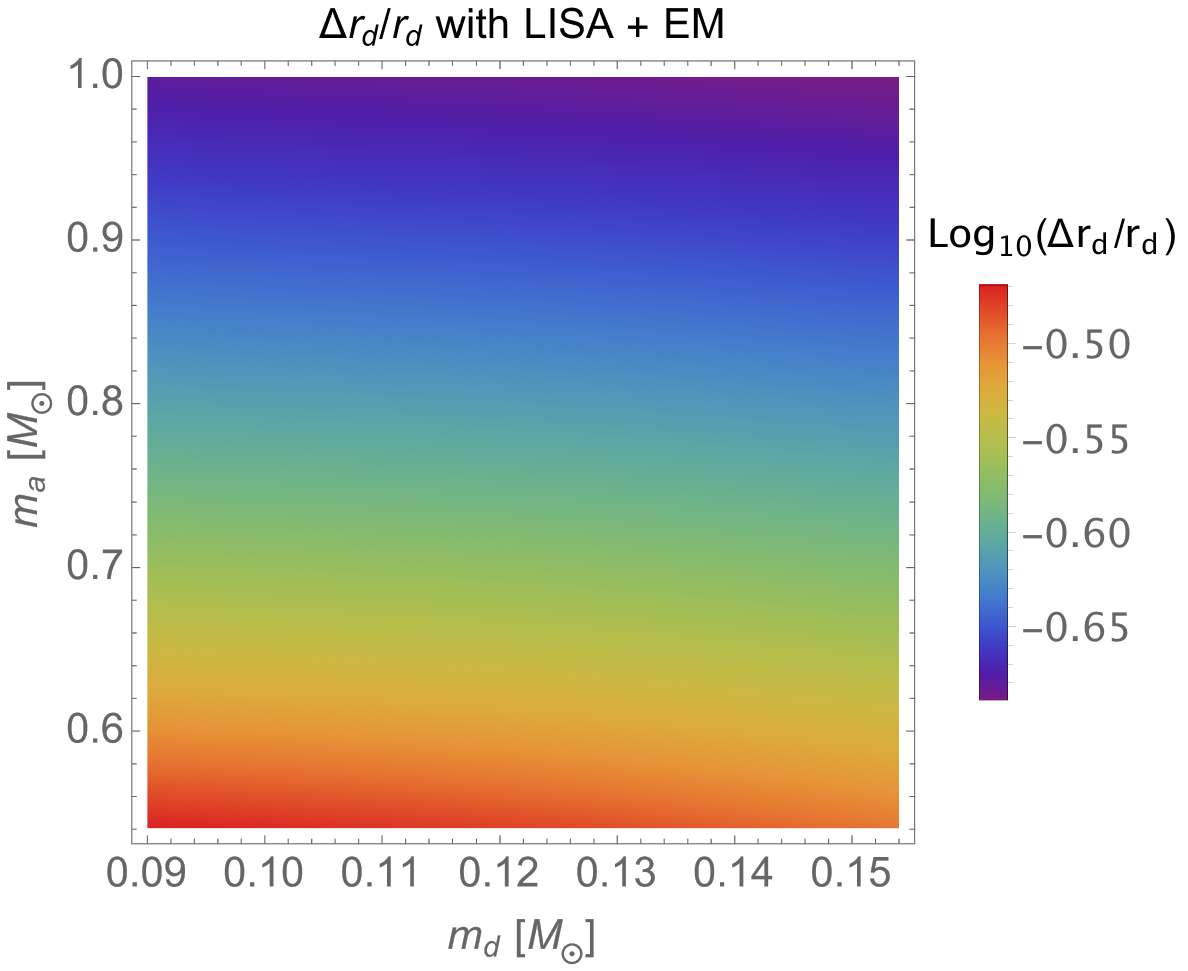} 
\caption{\emph{Top row:} Fractional error on the individual masses for non-degenerate DWD systems as determined via Fisher analysis using \textit{LISA} with electromagnetic counterparts. \emph{Bottom row, left to right:} Fractional error on $\eta_{\rm d}$ and $r_{\rm d}$ given by the same Fisher analysis. Fiducial values for $\eta_{\rm d}$ and $r_{\rm d}$ were obtained from the MESA mass-radius relation given in Fig.~\ref{fig:MESA}. As in Fig.~\ref{fig:6_params}, the accretion torque term was zero in these plots.
}
\label{fig:5_params}
\end{figure*}


Figure~\ref{fig:5_params} shows the measurement uncertainties calculated for the parameters $m_{\rm d}$, $m_{\rm a}$, $\eta_{\rm d},$ and $r_{\rm d},$ as determined via Fisher analysis including a prior on $D$ corresponding to a 40\% error on the parameter (i.e., $\sigma_D=0.4\times 8$~kpc). 
We see that if we have an independent observation of $D$, we can anticipate being able to constrain $m_{\rm d}$ and $m_{\rm a}$ (if $m_{\rm a}$ is sufficiently large for the latter), which we were unable to do without the complementary measurement of $D$. Although the measurability of $\eta_{\rm d}$ does not change significantly, the measurability of $r_{\rm d}$ is considerably enhanced when we perform the Fisher analysis with a prior on the luminosity distance. For measurement errors on $D$ itself, the analysis discussed here simply returns the prior we impose, so we do not gain any constraint on the luminosity distance.

Once again, we find that the error (without correlations) on $r_{\rm d}$ and $\eta_{\rm d}$ scales with SNR$\times (df/dr_{\rm d})\times r_{\rm d}$ and SNR$\times (d\dot f/d\eta_{\rm d})\times \eta_{\rm d}$, respectively. The errors on $m_{\rm d}$ and $m_{\rm a}$ do not follow such a simple scaling because of the priors that we impose on these parameters. Instead, we find that the error on $m_{\rm a}$ is mainly dominated by the prior, with a slight improvement that comes from the amplitude. On the other hand, the error on $m_{\rm d}$ is determined both from the amplitude and phase.

For the results shown in both Secs.~\ref{no GAIA} and~\ref{GAIA}, we note that the fractional errors on $m_{\rm d}$, $m_{\rm a},$ and $r_{\rm d}$ do not change significantly when we use the MESA model versus Eggleton's cold-temperature mass-radius relation for fiducial values of $r_{\rm d}$ and $\eta_{\rm d}$.\footnote{We would use Eggleton's mass-radius relation to perform parameter estimation on DWDs in a much later stage of evolution. We note that \textit{LISA} is less likely to be able to observe DWDs in this late stage, due to the significantly lower GW strain there (see Fig.~\ref{fig:strain}).} On the other hand, $\Delta \eta_{\rm d}/\eta_{\rm d}$ decreases significantly when we use Eggleton's mass-radius relation instead of a MESA model for fiducial values. This is because $d \dot{f}/d \eta_{\rm d}$ scales with $r_{\rm d}^{-11/2}$ (see Eqs.~\eqref{fnew} and~\eqref{fdot_total}, with $r_{\rm d}=r_{\rm L} a$). Evaluating this derivative at the smaller fiducial radius values given by the cold-temperature mass-radius relation (see Fig.~\ref{fig:MESA}) causes the Fisher matrix component for $\eta_{\rm d}$, which is determined by $d \dot{f}/d \eta_{\rm d}$, to be larger, leading to a smaller error (from inverting the Fisher matrix). 

Lastly, we note that to simplify calculations in both Secs.~\ref{no GAIA} and~\ref{GAIA}, we chose to average over the inclination, sky location, and polarization angles that appear in the general gravitational waveform of binaries detected by \textit{LISA} \citep{Cornish:2018dyw}; these angles do not appear in the equations we present in Sec.~\ref{waveform}. When we include these angles both in the waveform and as parameters in our Fisher matrix for several selected DWD systems, 
we find that the results are not noticeably affected; at most, the fractional error on $\eta_{\rm d}$ is increased a bit when we treat all the angles as unknown, but not by more than order unity for any of the cases we check.

\subsection{Application to ZTF J0127+5258}\label{ZTFJ0127+5258}

The discovery of ZTF J0127+5258 was very recently reported~\citep{Burdge:2023mtw}. This binary system, which has an orbital period of 13.7 minutes, is the first accreting verification DWD system for \textit{LISA} with a loud enough SNR and luminous donor. The binary is estimated to be at a distance of $3.5^{+1.7}_{-1.5}$kpc with a donor mass of either 0.19$\pm0.03M_\odot$ or 0.31$\pm0.11M_\odot$ and accretor mass of either 0.75$\pm0.06M_\odot$ or 0.87$\pm0.11M_\odot$, depending on the mass transfer rate.  Moreover, ZTF J0127+5258 is believed to be a DWD system in the pre-period minimum stage, which would confirm the presence of the DWDs we study within the 8kpc distance we have been considering.

We now show the results of our Fisher analysis method when we apply it to the astrophysical parameters of ZTF J0127+5258. Using the central values of measurements for the masses mentioned in the previous paragraph as fiducial values, along with fiducial $\eta_{\rm d}=1$\footnote{We choose some small, positive numbers for fiducial $\eta_{\rm d}$ to reflect lingering hydrogen on the donor of ZTF J0127+5258. While $\Delta \eta_{\rm d}$ depends significantly on what we choose for $F$ and fiducial $\eta_{\rm d}$, the other parameters' errors are agnostic to what we use for these values.} and $D=3.5$kpc, our Fisher analysis returns the measurement uncertainties compiled in Table~\ref{table:5_params}. Since we have a measurement of $D$ from ZTF, we perform a Fisher analysis with the 40\% error prior on $\sigma_D$.
The resultant calculations of statistical error due to detector noise are relatively small, meaning our constraints of these parameters from GW measurements are likely to be quite strong. We note that with \textit{LISA}, we can estimate the mass transfer rate from observations. However, the errors on $\dot{m}_d$ from \textit{LISA}'s observations are significant; if we propagate the errors due to $m_{\rm d}$, $m_{\rm a}$, $\eta_{\rm d}$, and $r_{\rm d}$ as listed in Table~\ref{table:5_params}, the propagated error on $\dot{m}_d$ overlaps with the standard deviation in mass transfer rate given by~\cite{Burdge:2023mtw} ($\log(\dot{M}/(M_\odot \mathrm{yr}^{-1}))=0.5$). Moreover, the errors we calculate for the individual parameters $m_{\rm d}$, $m_{\rm a}$, and $r_{\rm d}$ overlap with the errors from electromagnetic observations, which also suggests that we will not be able to use \textit{LISA}'s measurements to identify which of the mass transfer priors reported in~\cite{Burdge:2023mtw} is more accurate.

\begin{table}
\centering
\begin{tabular}{||c c c||} 
 \hline
& $\mathcal{N}(-8.5,0.5)$ & $\mathcal{N}(-7.3,0.5)$ \\ [1ex] 
 \hline
 $\Delta m_{\rm d}/m_{\rm d}$ & 0.7025 & 0.5198 \\ 
 $\Delta m_{\rm a}/m_{\rm a}$ & 0.8581  & 0.6586   \\
 $\Delta \eta_{\rm d}/
\eta_{\rm d}$ & 1.2790  & 0.5357  \\
 $\Delta r_{\rm d}/r_{\rm d}$ & 0.2407  & 0.1890  \\ [1ex] 
 \hline
\end{tabular}
\caption{Measurement uncertainties for astrophysical parameters of ZTF J0127+5258 calculated via our Fisher analysis. The two columns correspond to two sets of fiducial parameter values obtained from different priors of the mass transfer rate, assuming a normal distribution. The priors are denoted by $\mathcal{N}(a,b)$, with $a$ being the center value and $b$ the standard deviation in units of $M_\odot \mathrm{yr}^{-1}$. See \citet{Burdge:2023mtw} for details.}
\label{table:5_params}
\end{table}

\section{Conclusions}
\label{sec:conclusion}
We parameterize the GWs that we expect \textit{LISA} to detect from accreting DWD systems in terms of the parameters $\theta^i=(\phi_0,m_{\rm d},m_{\rm a},\eta_{\rm d},r_{\rm d},D)$. We perform a Fisher analysis on the parameterized waveform, imposing Gaussian priors on the individual masses based on the properties of the DWDs we expect to be generating the GWs. We find from our Fisher analysis that if we can obtain simultaneous, independent measurements of $D$ from a separate detector like Gaia, then we are likely to be able to constrain not only the individual masses, $m_{\rm d}$ and $m_{\rm a}$, but also $r_{\rm d}$. However, if we use only \textit{LISA}, lacking an independent measurement of $D$, then although our Fisher analysis still reveals reasonable measurability of $r_{\rm d}$, we lose our ability to constrain the individual masses. Finally, our parameter inference results suggest that we will be able to constrain astrophysical parameters of ZTF J0127+5258 from \textit{LISA}'s observations of the binary.

We note that, while the errors on other parameters are mostly dominated by the mass priors, 
it may be possible to constrain $\eta_{\rm d}$ somewhat better for a closer accreting DWD system ($D<8$ kpc) with a larger SNR. Importantly, an ability to constrain $\eta_{\rm d}$ might
be useful in distinguishing between the two scenarios of a hydrogen-rich donor ($\eta_{\rm d}\gtrsim1$) versus a cold, degenerate donor ($\eta_{\rm d}\approx-1/3$). To see this, we note that the errors obtained by Fisher analysis are interpreted as the standard deviation, $\sigma$, of a normal distribution centered at the best fit parameter. Given a fiducial $\eta_{\rm {d,fid}}$, we can therefore use our results to investigate the probability that $\eta_{\rm d}$ is in some range $\eta_{\rm {d,min}}-\eta_{\rm {d,max}}$, i.e.,
\begin{equation}
    \int_{\eta_{\rm {d,min}}}^{\eta_{\rm {d,max}}} P(\eta_{\rm d})  d\eta_{\rm d},
\end{equation}
where $P(\eta_{\rm d})=\mathcal{N}(\eta_{\rm {d,fid}},\sigma)$. In particular, for a given $\eta_{\rm {d,fid}}$, we can integrate the distribution from $-\infty$ to $-1/3$ to reveal the probability that the donor WD is in the late (T=0) stage of evolution with $\eta_{\rm d}<-1/3$. Repeating the above prescription with $r_{\rm d}$ instead of $\eta_{\rm d}$ would similarly allow us to distinguish between a finite temperature WD (with a larger radius) and a cold WD (with a smaller radius). Collectively, the distributions for $\eta_{\rm d}$ and $r_{\rm d}$ obtained from our Fisher analysis could lend insight into what stage of evolution the DWDs are in when we observe them with \textit{LISA}. We leave this for a future extension of this study.

There are a few additional avenues for future work to improve the current analysis. First, we have only used one evolutionary model of a hydrogen-rich donor to estimate errors on the physical parameters of all DWDs with such donors that \textit{LISA} will observe. To test the robustness of this analysis, one should study how much $\eta_{\rm d}$ can vary as a function of $m_{\rm d}$ between different hydrogen-rich donors, and see whether this variability affects our parameter inference. It would also be interesting to confirm and generalize our findings using a full Bayesian analysis~\citep{Cornish:2007if}. Some intrinsic assumptions in the Fisher approach, like having a high SNR, may be invalid and may result in substantial differences in the parameter constraints from the full Bayesian approach. Indeed, because we find SNR as low as $\sim 7$ in some regions of parameter space, it would be important to verify our findings with a more robust Bayesian analysis. Moreover, whereas a Fisher analysis can only provide Gaussian posterior distributions of the parameters, a Bayesian treatment provides the full posterior distributions, which can in general be non-Gaussian. One can also include the effect of tidal synchronization torque, as done in~\cite{Biscoveanu_2023,2015ApJ...806...76K,2017ApJ...846...95K}.

\section*{Data Availability}
The data used in this article will be shared on reasonable request to the corresponding authors.

\section*{Acknowledgements}

We thank Emanuele Berti for bringing the discovery of ZTF J0127+5258 to our attention. We all thank the support by NASA Grant No.80NSSC20K0523. S. Y. would also like to thank the UVA Harrison Undergraduate Research Award and the Virginia Space Grant Consortium (VSGC) Undergraduate Research Scholarship Program.



\bibliographystyle{mnras}
\bibliography{example} 




\appendix

\section{Corroboration of the $m_{\rm d}-f$ relation for negatively chirping DWD systems}\label{fgw-md}

\begin{figure}
\centering\includegraphics[width=0.48\textwidth]{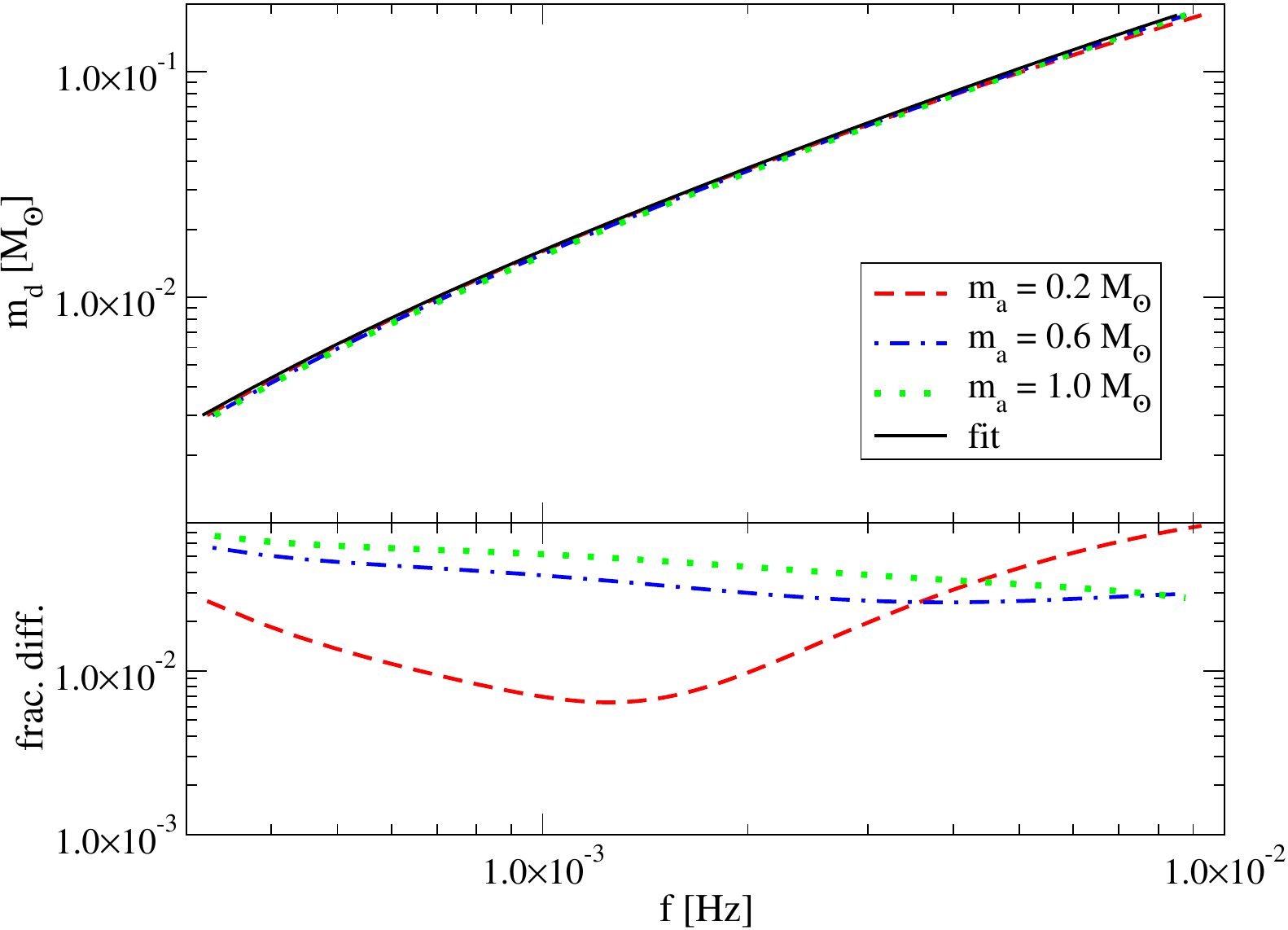} 
\caption{\emph{Top:} Our evolutionary tracks of $m_{\rm d}$ vs. $f$ for low-mass degenerate donors and various $m_{\rm a}$, as well as the analytic fit for similar DWD systems studied in~\citet{Breivik:2017jip}. \emph{Bottom:} Fractional difference between each $f$-$m_{\rm d}$ relation and the fit.}
\label{fig:breivik_fit}
\end{figure}

In a previous work,~\cite{Breivik:2017jip} report nearly identical evolutionary tracks of $m_{\rm d}$ versus $f$ in their simulations of several thousand negatively chirping DWDs containing low-mass helium core donors. They fit the relation between $m_{\rm d}$ and $f$ to a fourth-order polynomial shown in the top panel of Fig.~\ref{fig:breivik_fit}. In the same figure, we plot tracks of $m_{\rm d}$ versus $f$ calculated via Eq.~\eqref{fnew}, using Eggleton's cold-temperature mass-radius relation to obtain donor radius values~\citep{1988ApJ...332..193V}. It is evident that despite the accretor mass appearing in the total mass in Eq.~\eqref{fnew}, the evolution of $m_{\rm d}$ with $f$ is largely insensitive to the mass of the accreting companion in the negatively chirping regime; the fractional difference between our relations with different $m_{\rm a}$ values and the analytic fit is less than 8\% (see the bottom panel of Fig.~\ref{fig:breivik_fit}). For DWDs containing low-mass degenerate donors, our analytic calculation of $f$ agrees excellently with the analytic fit in Eq. (1) of~\cite{Breivik:2017jip}. 

\section{Further Analysis on Parameter Estimation Results}
\label{scaling}
\begin{figure*}
\centering\includegraphics[width=0.83\textwidth]{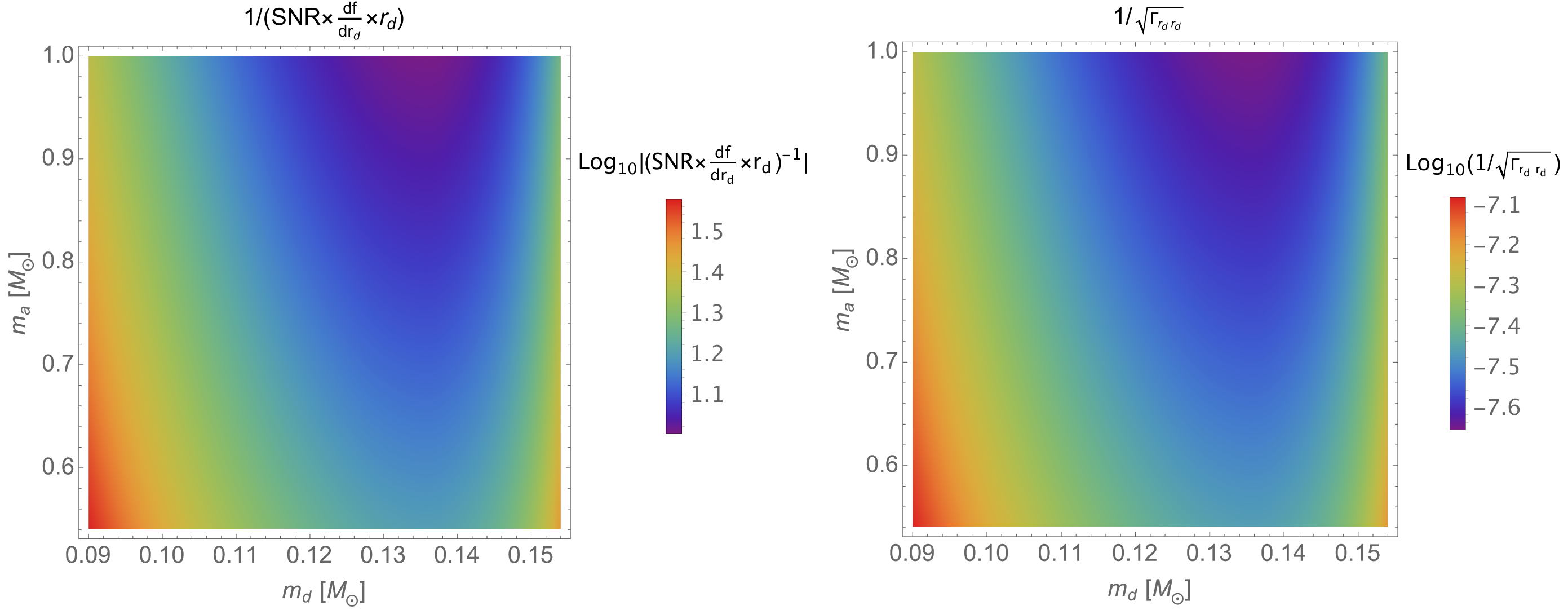} 
\centering\includegraphics[width=0.82\textwidth]{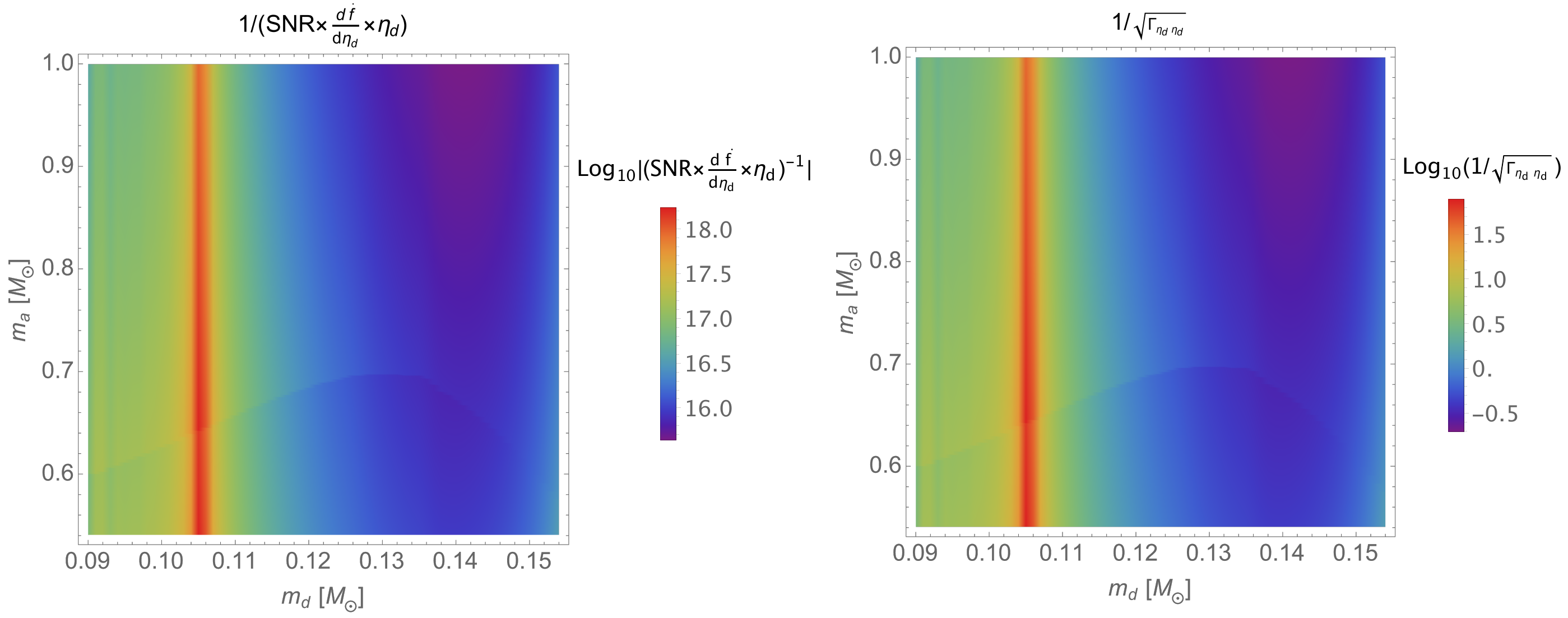} 
\centering\includegraphics[width=0.83\textwidth]{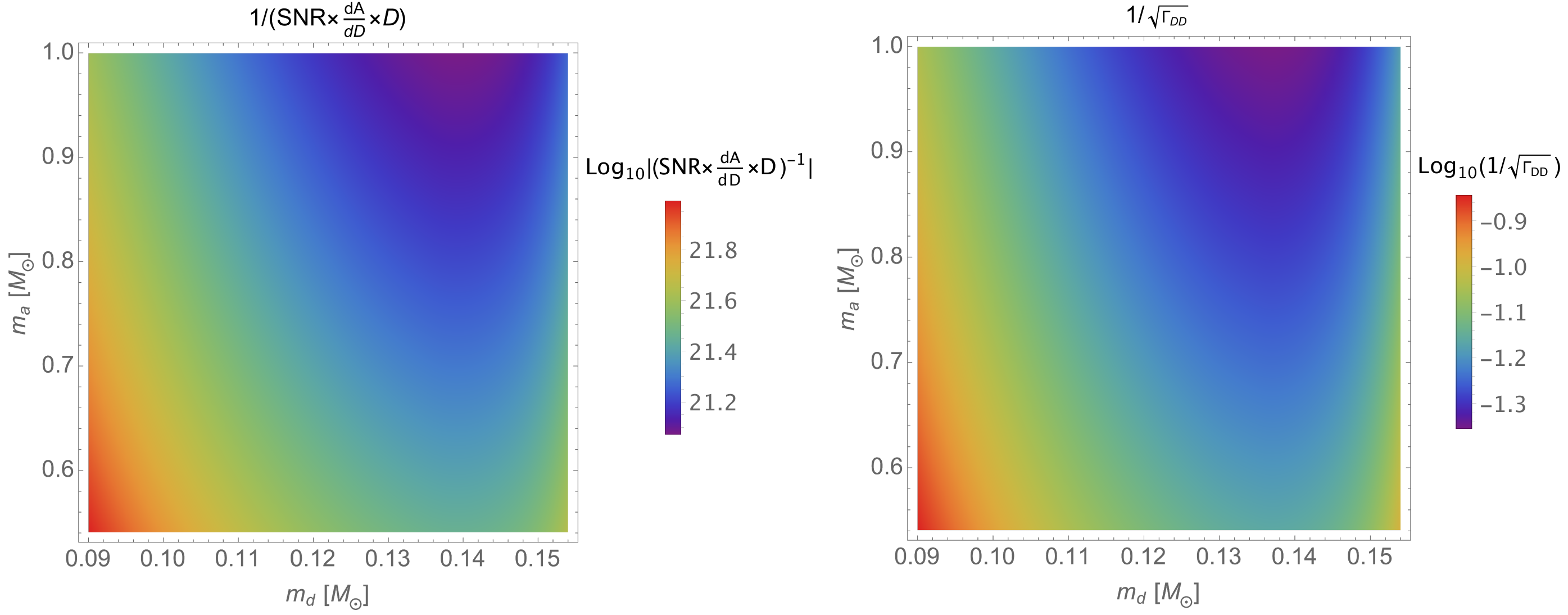} 
\caption{Plots illustrating how errors on $r_{\rm d}$, $\eta_{\rm d}$, and $D$  (right) depend most heavily on the SNR and $f, \dot{f},$ and $A$, respectively (left). The quantities on the right increase or decrease in the same manner as the quantities on the left. This suggests that our ability to measure WD parameters $(r_{\rm d}, \eta_{\rm d}, D)$ is dominated by the SNR and how well we measure raw parameters $(f, \dot{f}, A)$, respectively.
}
\label{fig:app_scalings}
\end{figure*}

In Secs.~\ref{no GAIA} and \ref{GAIA}, we mention the inverse scaling of the error on $r_{\rm d}$, $\eta_{\rm d}$, and $D$ with the SNR times $df/dr_{\rm d}$, $d\dot f/d\eta_{\rm d}$, and $dA/dD$ respectively. 
To illustrate this point, we plot in Fig.~\ref{fig:app_scalings}  the error on each of these parameters without correlations alongside the inverse of the partial derivative of the waveform with which the error scales. As before, we obtain the error without correlations on a parameter $\theta^i$ by computing $1/\sqrt{\Gamma_{ii}}$, as opposed to taking the $i^{\rm{th}}$ diagonal component of the full inverted Fisher matrix. This quantity reveals only the strength of the dependence of the gravitational waveform on the $i^{\rm{th}}$ parameter, whereas the full errors plotted in Fig.~\ref{fig:6_params} also take into account the limitations to measuring the parameter due to its interdependence on the other parameters. Indeed, we see that the errors on $r_d$ and $D$ especially are affected by the measurability of the individual masses, so that the plots for these parameters in Fig.~\ref{fig:app_scalings} are qualitatively quite different from the corresponding plots in Fig.~\ref{fig:6_params}. 

In Fig.~\ref{fig:app_scalings}, we see that the plots in the left column scale with plots on the right. That is, the magnitude of error (without correlations) on parameters $r_{\rm d},\eta_{\rm d},$ and $D$ is reliant on the SNR and raw model parameters $f, \dot{f},$ and $A$, respectively.  This suggests that out of the four raw model parameters in the waveform, $(\phi_0,A,f,\dot{f})$,
$r_{\rm d}$ is largely determined by $f$, $\eta_{\rm d}$ is determined by $\dot f$, and $D$ is determined by $A$. The latter two statements are not surprising; $\eta_{\rm d}$ and $D$ only appear in the waveform through $\dot f$ and $A$, respectively, so we would not expect the error on these parameters to be affected by anything else (other than the SNR). The clear scaling of $r_{\rm d}$ with $f$ is slightly less trivial since $r_{\rm d}$ technically appears in all three pieces of the waveform, $A$, $f$, and $\dot f$. However, since $r_{\rm d}$ only enters the amplitude and $\dot f$ through $f$, we should not ultimately be surprised that $r_{\rm d}$ scales most strongly with just the GW frequency (and SNR).




\bsp	
\label{lastpage}
\end{document}